\documentclass[preprint]{aastex}

\slugcomment{Accepted for publication in {\it ApJ} (Volume 593, 10 August 
2003)}

\shorttitle{{\it RXTE}, {\it ROSAT} and {\it ASCA} Observations 
of G347.3-0.5}
\shortauthors{Pannuti et al.}

\begin{document}


\title{{\it RXTE}, {\it ROSAT} and {\it ASCA} Observations of G347.3$-$0.5\\ 
    (RX J1713.7$-$3946): Probing Cosmic-Ray Acceleration\\ 
    by a Galactic Shell-Type Supernova Remnant}


\author{Thomas G. Pannuti\altaffilmark{1}, Glenn E. 
Allen\altaffilmark{1}, John C. Houck\altaffilmark{1} and
Steven J. Sturner\altaffilmark{2,3}}


\altaffiltext{1}{MIT Center for Space Research, 77 Massachusetts Avenue,
Cambridge, MA 02139; tpannuti@space.mit.edu, gea@space.mit.edu,
houck@space.mit.edu}
\altaffiltext{2}{NASA Goddard Space Flight Center, Code 661, Greenbelt, MD
20771; sturner@swati.gsfc.nasa.gov}  
\altaffiltext{3}{Universities Space Research Association (USRA), 7501
Forbes Blvd., Suite 206, Seabrook, MD 20706-2253}


\begin{abstract}
We present an analysis of the X-ray spectrum of the Galactic shell-type 
SNR G347.3$-$0.5 (RX J1713.7$-$3946). This SNR is a member of a 
growing class of SNRs which are dynamically young, shell-type sources that 
emit non-thermal X-rays from specific regions on their outer shells. 
By performing a joint spectral analysis of data from observations made 
of G347.3$-$0.5 using the $\it{ROSAT}$~PSPC, the $\it{ASCA}$~GIS and the 
$\it{RXTE}$~PCA, we have fit the spectra of particular regions of this 
SNR (including the bright northwestern and southwestern rims, the 
northeast rim and the interior diffuse emission) over the approximate
energy range of 0.5 through 30 keV. We find that fits to the spectra of 
this SNR over this energy range using the $\it{SRCUT}$ model were superior 
to a simple power law model or the $\it{SRESC}$ model. We find that the 
inclusion of a thermal model with the $\it{SRCUT}$ model helps to improve 
the fit to the observed X-ray spectrum: this represents the first detection 
of thermal X-ray emission from G347.3$-$0.5. Thermal emission appears to be 
more clearly associated with the diffuse emission in the interior of the SNR 
than with the bright X-ray emitting rims. A weak emission feature seen 
near 6.4 keV in the $\it{RXTE}$~PCA spectrum most likely 
originates from diffuse X-ray emission from the surrounding Galactic Ridge 
rather than from G347.3$-$0.5 itself. We have analyzed our $\it{RXTE}$~PCA 
data to search for pulsations from a recently discovered radio pulsar 
(PSR J1713$-$3949) which may be associated with G347.3$-$0.5, and we do
not detect any X-ray pulsations at the measured radio period of 392 ms. 
Using the best-fit parameters obtained 
from the $\it{SRCUT}$ model, we estimate the maximum energy of 
cosmic-ray electrons accelerated by the rims of G347.3$-$0.5 to be 19-25 
TeV (assuming a magnetic field strength of $B$ = 10$\mu$G), consistent 
with the results of \citet{ESG01}. We present a broadband (radio to 
$\gamma$-ray) photon energy-flux spectrum for the northwestern rim of 
G347.3$-$0.5, where we have fit the spectrum using a more sophisticated 
synchrotron-inverse 
Compton model with a variable magnetic field strength. Our fit
derived from this model yields a maximum energy of only 8.8$_{-3.4}^{+4.1}$ 
TeV for the accelerated cosmic-ray electrons and a much greater magnetic 
field strength of 150$_{-80}^{+250}$ $\mu$G: however, our derived
ratio of volumes for TeV emission and X-ray emission based on this fit -- 
$V$$_{TeV}$/$V$$_{X-ray}$ $\approx$ 1000 -- is too large to be physically
acceptable. We argue that neither non-thermal bremsstrahlung nor 
neutral pion paritcle decay can adequately explain the TeV emission from 
this rim, and therefore
the physical process responsible for this emission at this site is 
currently uncertain. Finally, we compare the 
gross properties of G347.3$-$0.5 with other SNRs known to possess X-ray 
spectra dominated by non-thermal emission.
\end{abstract}




\section{Introduction}

Supernova remnants (SNRs) are considered the leading candidates 
for the accelerators of cosmic-rays to at least the ``knee" of
the cosmic-ray energy spectrum ($E$ $\approx$ 3000 TeV). While the
relationship between SNRs and cosmic-ray acceleration has received 
extensive attention in the literature 
\citep{Bell78a, Bell78b, Axford81, BE87, Jones01, M01}, 
most of this discussion has been 
theoretical, with only a limited amount of observational data to consider. 
One major unresolved issue related to cosmic-ray acceleration by
SNRs is the maximum energy $E$$_{cutoff}$ attained by cosmic-ray particles 
accelerated by the shock front: for samples of young shell-type 
X-ray emitting SNRs located in the
Galaxy and the Large Magellanic Cloud, \citet{RK99} and \citet{HR01}
estimated values for $E$$_{cutoff}$ of $\leq$ 200 TeV and $\leq$ 80 TeV,
respectively, both well below the cosmic-ray energy spectrum ``knee."
Progress on this issue has been limited due to the lack of both X-ray
and TeV observations of SNRs with the angular and spectral resolution needed 
to probe the acceleration process. 
\par
Recently, however, high spatial resolution observations of SNRs at X-ray
energies have become feasible for the first time using such X-ray 
observatories as $\it{ROSAT}$, $\it{ASCA}$, $\it{Chandra}$ and $\it{XMM}$. 
These high resolution observations yield spatially-resolved
spectroscopy of X-ray emitting features within the SNR, particularly 
the X-ray luminous rims that are associated with the
expanding shock fronts of SNRs. By performing a detailed spectral 
analysis of X-ray emission from the rims of SNRs, corresponding to shock
fronts where interstellar or circumstellar material is being swept up, the 
cosmic-ray
acceleration process may be probed. Initial insights on cosmic-ray
acceleration by SNRs have been gleaned from $\it{ROSAT}$, $\it{ASCA}$
and $\it{RXTE}$ observations of the high-energy X-ray emission from such
SNRs as Cas A and SN 1006 \citep{W96, A97, Keohane98, Dyer01, A01}, 
but a thorough multi-wavelength study of SNRs is required in order to 
obtain a detailed understanding of the cosmic-ray acceleration process.  
\par
In this paper, we present our analysis of the non-thermal 
X-ray emission from the Galactic SNR G347.3$-$0.5 (also known 
as RX J1713.7$-$3949). First discovered during the R\"{o}ntgensatellit 
($\it{ROSAT}$) All Sky Survey \citep{PA96}, this SNR has 
attracted a considerable amount of interest from the astronomical
community. G347.3$-$0.5 is one of the brightest Galactic X-ray 
SNRs known with an X-ray flux density of 4.4 $\times$ 10$^{-10}$ erg 
cm$^{-2}$ s$^{-1}$ \citep{PA96}. Subsequent studies of this SNR based
on $\it{ROSAT}$ and $\it{ASCA}$ observations 
\citep{K97,S99} revealed that most of its X-ray emission is non-thermal
and most likely produced by the synchrotron process. Unlike the X-ray
spectra typically seen from most other SNRs, no thermal line emission was
detected from any portion of G347.3$-$0.5 by these previous studies. 
The detection of non-thermal X-ray emission from this SNR has made
G347.3$-$0.5 one of the prime examples of an SNR acting as an accelerator
of cosmic-ray particles. In addition, \citet{M00} detected emission from 
TeV photons located on the northwestern rim of G347.3$-$0.5: those authors
argued that these photons were produced by the inverse-Compton scattering
process, with cosmic microwave background photons scattering off 
high-energy cosmic-ray electrons accelerated along the shock front of
this SNR. This result makes G347.3$-$0.5 only the third shell-type 
SNR besides SN 1006 \citep{K95} and Cas A \citep{Aharonian01} 
reported to produce TeV emission. 
A detailed analysis of the evolutionary state of G347.3$-$0.5 using radio, 
X-ray and TeV observations of the bright northwestern rim was given by 
\citet{ESG01}. Those authors also estimated the maximum energy of 
cosmic-ray protons accelerated by G347.3$-$0.5 to be approximately 20 TeV. 
\citet{Butt01} argued that this SNR is accelerating 
cosmic-ray nuclei as well as cosmic-ray electrons based on positional 
coincidence between the northeastern rim of G347.3$-$0.5 and the 
$\gamma$-ray source 3EG J1714$-$3857 \citep{H99}, which is positionally 
coincident with a molecular cloud: this cloud is assumed to be interacting 
with G347.3$-$0.5. Cosmic ray acceleration by this SNR was also
discussed by \citet{Enomoto02}, who presented a TeV spectrum for the 
northwestern rim of G347.3$-$0.5 and argued that this spectrum is best fit by 
the decay of neutral pion ($\pi$$^0$) particles rather than an inverse
Compton spectrum of upscattered cosmic background photons. These pions
are presumed to be produced by the collision of cosmic-ray protons
(assumed to be accelerated by G347.3$-$0.5) with protons in the molecular
cloud, as described by \citet{Enomoto02}. This result is in part based on 
the assumption that 3EG J1714$-$3857 is associated with G347.4-0.5 and,
if correct, would represent significant observational evidence for cosmic-ray 
proton acceleration by an SNR \citep{Aharonian02}. However, \citet{RP02} 
and \citet{Butt02} have disputed the association with the $\gamma$-ray
source and instead claim that, while G347.3$-$0.5 is very likely an 
accelerator of cosmic-ray electrons, no convincing evidence yet exists 
that it accelerates hadrons as well. Other recent work on G347.3$-$0.5
includes modeling of the spectrum of AX J1714.1$-$3912, a hard X-ray
source associated with 3EG J1714$-$3857 \citep{Uchiyama02}, the discovery 
of fine-structure
filaments associated with the northwestern rim of G347.3$-$0.5 
\citep{Uchiyama03} and finally, a discussion of the
flux of high energy neutrinos from the northwestern rim of this SNR,
assuming that the TeV spectrum is produced by the decay of neutral
pions, which yield neutrinos in the decay process \citep{AMH02}. In
this paper, we present a spatially-resolved broadband spectral
analysis of G347.3$-$0.5 in order
to analyze the high energy emission from this SNR and probe cosmic-ray
acceleration by this source. Like many Galactic SNRs, the distance to 
G347.3$-$0.5 and its age are not well known: early studies of this SNR 
suggested a low distance of about 1 kpc assuming  
G347.3$-$0.5 is in the Sedov stage of evolution 
\citep{PA96} or that it is associated with the historical supernova of 
393 A.D. \citep{W97}. In addition, \citet{K97} derived a distance of
1 kpc by comparing their measurement of the column density
toward this source with the total column density in the direction of
the Galactic Center. Such a low distance suggests that G347.3$-$0.5 is
a young SNR with an age of about 2000 years. However, a subsequent 
analysis by \citet{S99} instead favored a larger distance of 6 kpc 
based on a postulated association between G347.3$-$0.5, an HII region and a
complex of three molecular clouds: such a distance would imply a much
greater age of about 10$^4$ years. We will adopt a distance to G347.3$-$0.5
of 6 kpc for this paper in order to be consistent with the analyses
presented by \citet{S99} and \citet{ESG01} as well as the putative
association between this SNR and the recently discovered pulsar 
J1713-3949, believed to lie at a distance of 5.0$\pm$0.2 kpc based
on measurements of its dispersion measure \citep{Crawford02}.
\par
The organization of this paper is as follows: in Section \ref{ObsSection},
we describe the observations of G347.3$-$0.5 using {\it ROSAT} (Section 
\ref{ROSATSubSection}), {\it ASCA} (Section \ref{ASCASubSection}) and 
{\it RXTE} (Section \ref{RXTESubSection}). While the {\it ROSAT} and
{\it ASCA} observations have been presented previously, the {\it RXTE}
observations are presented here for the first time. The spectral
fitting and analysis for the datasets from all of the observations
are described in Section \ref{SpectFitAnalysisSection}, including 
descriptions of the models used to fit the non-thermal and thermal 
components to the X-ray emission. The results of
this spectral fitting are presented in Section \ref{ResultsSection}:
we discuss in turn the maximum energies of electrons shock-accelerated
by G347.3$-$0.5 (Section \ref{MaxEnergySubSection}), our search for
thermal X-ray emission from different regions of this SNR (Section
\ref{ThermalSubSection}), and the nature of the weak emission line
found in the {\it RXTE}~PCA spectrum of this source 
(Section \ref{IronFeatureSubSection}). In our {\it RXTE}~PCA data, we
have searched for pulsed X-ray emission from the recently detected radio 
pulsar which is possibly associated with the discrete X-ray source seen
at the center of G347.3$-$0.5: we describe this search in Section 
\ref{PulsarSearchSubSection}. A study of the broad-band 
spectrum of the northwestern rim of G347.3$-$0.5 from radio to 
$\gamma$-ray wavelengths is presented in Section \ref{NWRimSection}.
Comparisons between G347.3$-$0.5 and two similar SNRs (SN 1006 and
G266.2$-$1.2) are presented in Section \ref{ComparisonSection}, and
lastly the conclusions are given in Section \ref{ConclusionsSection}.

\section{Observations and Data Reduction\label{ObsSection}}

In this Section we discuss the X-ray observations that were
used in this analysis, including the instruments used and the data
reduction processes. These observations include archived 
data from observations of G347.3$-$0.5 that were made using the Position
Sensitive Proprtion Counter (PSPCB, hereafter referred to as just PSPC)
of $\it{ROSAT}$ and the Gas Imaging Spectrometer (GIS) of $\it{ASCA}$.
We include our observations of this SNR using the Proportional Counting 
Array (PCA) of the Rossi X-ray Timing Explorer ($\it{RXTE}$). These data 
sample the X-ray emission from G347.3$-$0.5 over nearly two full decades 
of energy (0.5 keV to 30 keV). Details of the observations for each
instrument are provided in Table \ref{ObsTable}. 
\subsection{$\it{ROSAT}$ PSPC Observations and Data 
Reduction\label{ROSATSubSection}}
$\it{ROSAT}$~observed G347.3$-$0.5 at a location near the center of the
remnant using the PSPC, a multiple-wire proportional 
counter sensitive to photons that have energies between approximately 
0.1 and 2.0 keV. The energy resolution $\Delta$$E$/$E$ is 0.43 at 0.93 
keV and the field of view of the telescope is 2$^{\circ}$. For our
off-axis observations, the angular resolution is $\approx$ 50 arcseconds
(50\% encircled-photon radius), and the maximum on-axis effective area of 
the system is about 260 cm$^2$ \citep{P87}. 
\par
\begin{figure}
\figurenum{1}
\plotone{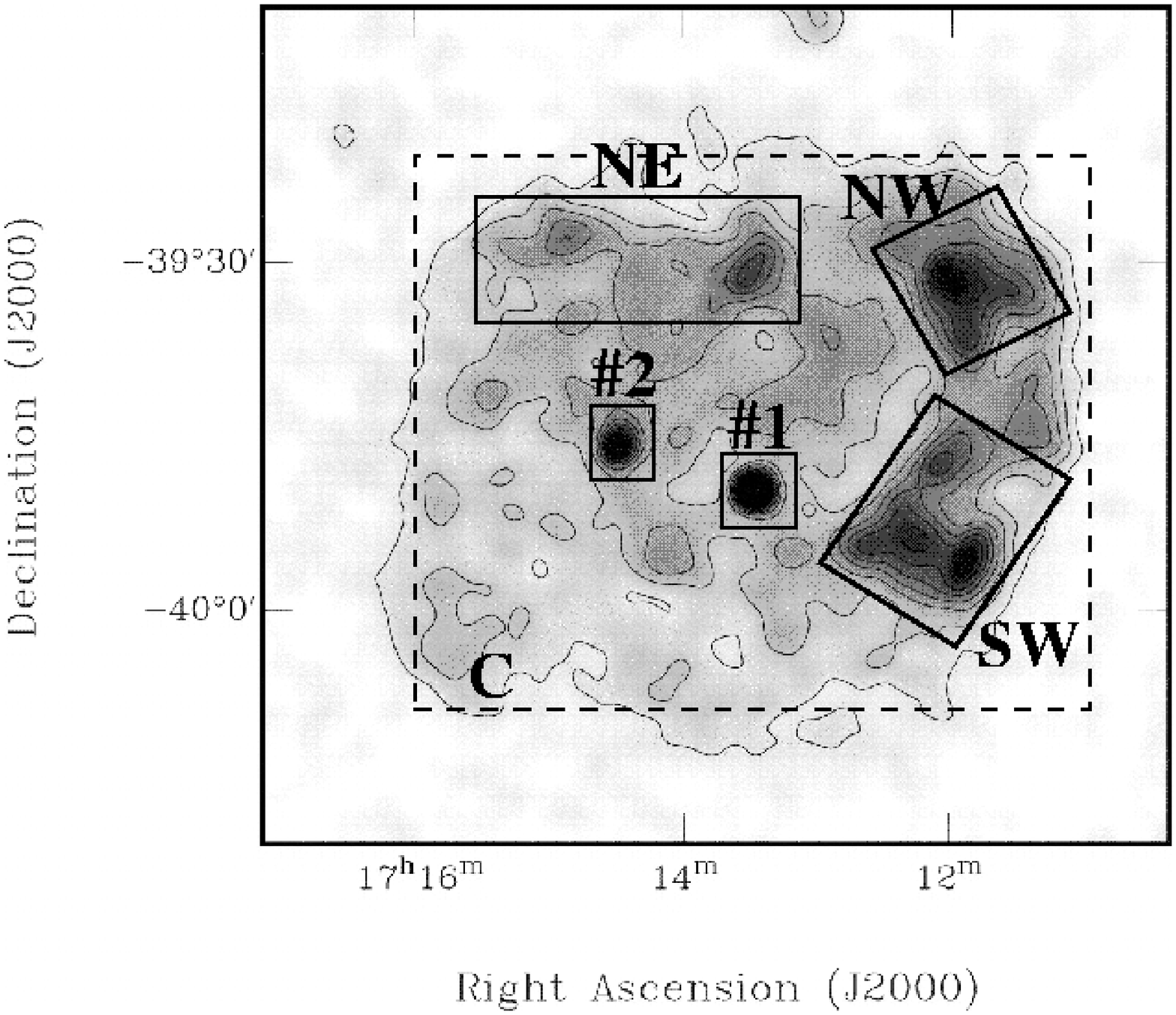}
\caption{{\it ROSAT}~PSPC image of G347.3$-$0.5 from \citet{S99}.
Note the X-ray luminous northwestern rim (a TeV source) as well as
the bright southwestern and northeastern rims. Contour levels start
at 1.44 counts per arcmin$^{-2}$ sec$^{-1}$ and increase upward in
steps of 1.15 counts per arcmin$^{-2}$ sec$^{-1}$. The eastern bright
source (labeled as ``\#2") is a foreground star, while the central source
(labeled as ``\#1" and also known as 1WGA J1713.4$-$3949) may be 
associated with the radio pulsar PSR J1713$-$3949. Discrete
regions of extracted spectra -- namely the northwestern rim (``NW"), 
the southwestern rim (``SW") and the northeastern rim (``NE") -- are
indicated by solid boxes. The central diffuse emission (``C") is indicated
by a dashed box: flux from the three rims and the two interior point
sources were excised from the flux from this region.\label{fig1}}
\end{figure}

Spectra were
constructed using the PSPC data for four spatially separate regions:
the luminous northwestern and southwestern rims, the northeastern
rim and the interior region. This final region was considered in
order to analyze diffuse emission from the interior of the SNR. Emission
from the two bright interior point sources -- 1WGA J1714.4$-$3945 and
1WGA J1713.4$-$3949 \citep{PA96,S99} -- were subtracted from this 
region before a spectrum was extracted. The former source appears
to be associated with a foreground star while the latter may be
the X-ray counterpart to a recently-discovered radio pulsar. This
radio pulsar may in turn be associated with G347.3$-$0.5: we will discuss 
this source in more detail in Section \ref{PulsarSearchSubSection}.
An image from the PSPC observation appears in Figure \ref{fig1} with
the positions of these regions indicated. Fits to the spectra extracted
from these regions will be described in Section 
\ref{JointSpectralFitSubSection}.
\subsection{$\it{ASCA}$ GIS Observations and Data 
Reduction\label{ASCASubSection}}
The instruments on $\it{ASCA}$ \citep{TIH94} observed locations on
the southwestern rim, the northwestern rim and the northeastern rim
of G347.3$-$0.5. While both the GIS and the Solid-State Imaging 
Spectrometer 
(SIS) aboard $\it{ASCA}$ collected data during these observations, the GIS 
has a larger field of view so only the data from the GIS were considered 
for the present study. The GIS consists of two units (denoted as GIS2 and 
GIS3) of imaging gas scintillation proportional counters with a sealed-off 
gas cell equipped with an imaging phototube: for our research, we only 
considered data collected by the GIS2 unit. The GIS is sensitive over the
energy range of approximately 0.7 - 10 keV and has an energy resolution of
$\Delta$$E$/$E$ = 0.078 at 6 keV \citep{O96, M96}. The field of view of 
the GIS is 22$\arcmin$~in radius and the maximum effective area for our 
off-axis observations is 94 cm$^2$ at 2.07 keV. The data from these
observations were reduced using standard $\it{ASCA}$ GIS 
reduction techniques as described in ``The $\it{ASCA}$~Data Reduction
Guide, Version 2.0." The reduction process filtered data for elevation
angle, stable pointing directions, the South Atlantic Anomaly and 
cut-off rigidity. The data were also screened for characteristics of the GIS 
internal background and screening based on event locations and rise time. The 
standard script ``ASCASCREEN" was run in order to accomplish much of this 
reduction. We extracted spectra for the northwestern rim, the southwestern 
rim and the northeast rim of the SNR using the same regions as in the 
$\it{ROSAT}$~PSPC spectral analysis. 
Fits to these spectra will be discussed in detail in
Section \ref{JointSpectralFitSubSection}.
\subsection{$\it{RXTE}$ PCA Observations and Data 
Reduction\label{RXTESubSection}}
G347.3$-$0.5 was observed by the $\it{RXTE}$~PCA with a pointing position
that was offset by 7.8\arcmin~from the nominal center of the SNR to
avoid contaminating flux from three nearby X-ray binaries (4U1708$-$408,
1RXS J170849.0$-$400910 and GPS1713$-$388) that appear close to 
G347.3$-$0.5 in the sky. Each of these X-ray binaries is located 
approximately 1$^{\circ}$ from the pointing position, so none of
them appeared within the field of view (1$^{\circ}$ FWHM) of the
RXTE {\it PCA}~observation. Therefore, effects of contaminating
flux from these X-ray binaries may be safely ignored. 
\par
The PCA is a spectrophotometer comprised of an array of five coaligned
proportional counter units that are mechanically collimated to have
a field of view of 1$^{\circ}$ FWHM. The array is sensitive to photons 
that have energies between approximately 2 and 60 keV. The energy 
resolution $\Delta$$E$/$E$ of the array is 0.18 at 6 keV, and
the maximum on-axis collecting area is about 6000 cm$^2$ at 9.72 keV.
The PCA data were screened to remove the time intervals
during which (1) one or more of the five proportional counter units
is off, (2) G347.3$-$0.5 is less than 10$^{\circ}$ above the limb of
the Earth, (3) the background model is not well-defined and (4) the
pointing direction of the detectors is more that 0$^{\circ}$.02 from  
the nominal pointing direction in either right ascension or declination.
After screening the data using these criteria, a final dataset was 
prepared. The PCA background spectrum for G347.3$-$0.5 was estimated
using the the FTOOL $\it{pcabackest}$\footnote{See
http://heasarc.gsfc.nasa.gov/docs/software/ftools/ftools\_menu.html
for more information about FTOOLS.}.
\par
\begin{figure}
\figurenum{2}
\plotone{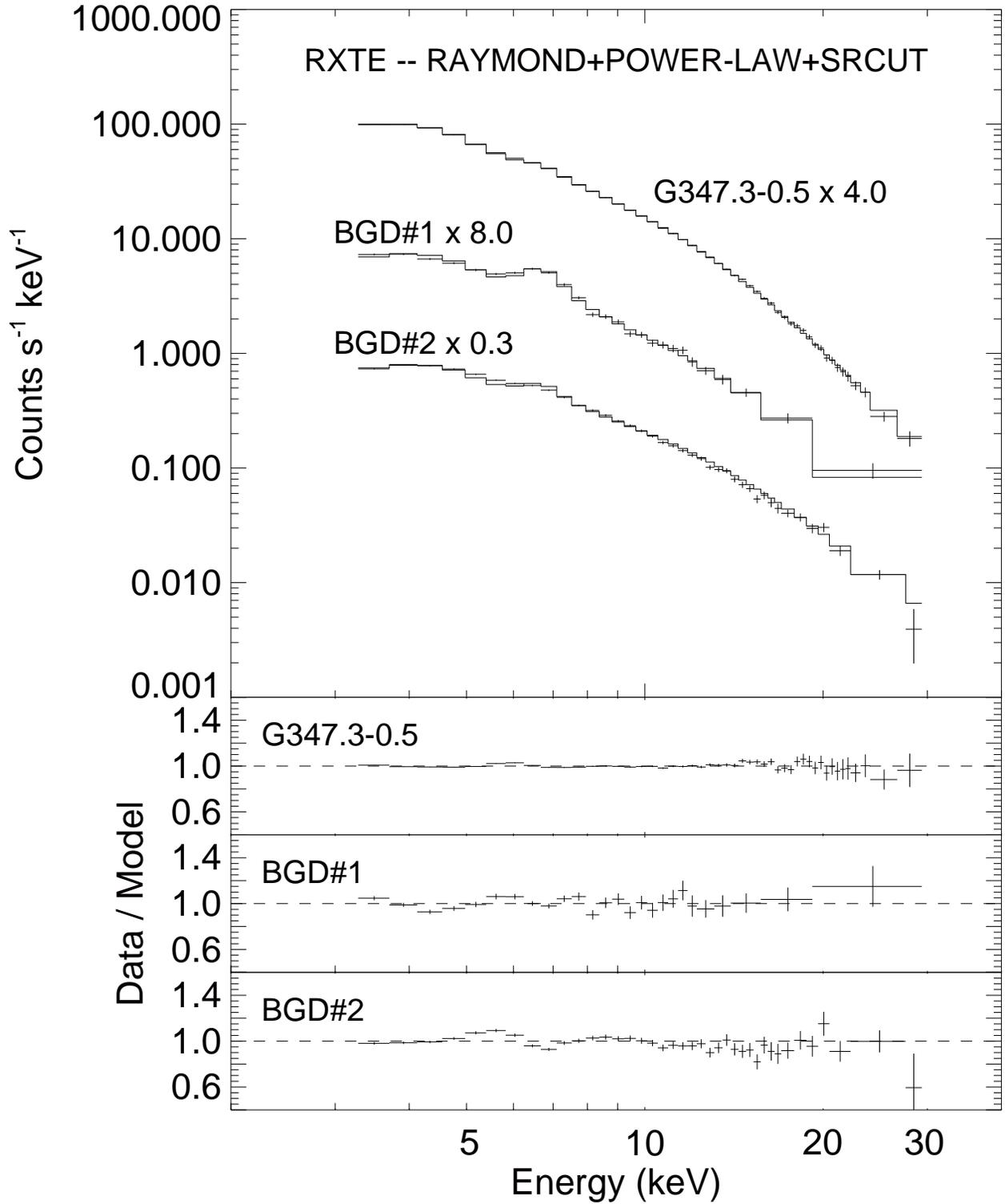}
\caption{Spectra from {\it RXTE} PCA observations of G347.3$-$0.5
and two background regions (labeled as ``BGD\#1" and ``BGD\#2")
using an SRCUT model combined with a Power Law+Raymond component
to fit the Galactic Ridge emission using model parameters derived
by \citet{VM98}.\label{fig2}}
\end{figure}

G347.3$-$0.5 lies toward the direction of the Galactic Ridge (GR), a region
of the Galaxy which features extensive diffuse X-ray emission as
described by \citet{VM98}. G347.3$-$0.5 falls within
the sub-region of the GR denoted as R1 by those authors:
for this sub-region, \citet{VM98} obtained a best-fit for the observed
X-ray emission using a two component model. The first component is
a Raymond-Smith plasma component \citep{RS77} -- denoted as 
$\it{RAYMOND}$ -- which describes the emission spectrum from a hot diffuse 
gas, including line emissions from several elements. Using this model,
with abundances frozen to solar, \citet{VM98} obtained 
a temperature of $\it{kT}$ = 2.9 keV for the thermal emission from this 
sub-region of the GR. The second component was a simple 
power law (denoted as $\it{POWER~LAW}$) which models the 
non-thermal emission from this sub-region: \citet{VM98} obtained a photon 
index of 1.8 for this component. Lastly, the derived column 
density for the spectral fit was $N$$_H$ = 1.8$\times$10$^{22}$ cm$^{-2}$.
To sample this background X-ray emission, two additional
observations were made with the PCA of fields adjacent to the field
observed in the pointed observation of G347.3$-$0.5: we denote these
fields as Background Region 1 and Background Region 2, respectively (see 
Table \ref{ObsTable}). We note that the approximate pointing
errors for all of the PCA observations are $\approx$8\arcsec.
Data from these two observations were reduced in the same way as
the data from
the pointed observation of G347.3$-$0.5. In Figure \ref{fig2}, we 
present spectra extracted for the on-source observation and both
of the background observations. We will return to a discussion of
X-ray emission from the GR in Section \ref{IronFeatureSubSection}.
\section{Spectral Fitting and Analysis\label{SpectFitAnalysisSection}}
\subsection{The $\it{RXTE}$ Spectrum of G347.3$-$0.5}
After reducing all of the data, we proceeded to perform spectral-fitting
analysis using the X-ray spectral-fitting package $\it{XSPEC}$\footnote{See
http://heasarc.gsfc.nasa.gov/docs/xanadu/xspec/index.html for
more information about XSPEC.}, first considering the $\it{RXTE}$ PCA 
data. Given that the X-ray emission from 
G347.3$-$0.5 is known to be almost entirely non-thermal 
and probably produced by electrons emitting
synchrotron radiation, we used three models for non-thermal X-ray
emission to fit the data: a simple power law, the $\it{SRESC}$ model
and the $\it{SRCUT}$ model. An emission feature is seen 
in the $\it{RXTE}$~PCA spectrum near 6.4 keV: to 
model this feature, we included a Gaussian component with each of the 
models used in the spectral fitting (see Section 
\ref{IronFeatureSubSection} for a more detailed discussion about
this apparent feature). In all cases, photoelectric absorption 
along the line of sight was modeled using the Wisconsin \citep{MM83} 
cross-sections and relative abundances of elements as described by
\citet{AE82}. 
\par
$\it{POWER~LAW}$: Earlier studies of the X-ray spectra 
associated with the bright rims of G347.3$-$0.5 by \citet{K97} and 
\citet{S99} revealed that the X-ray emission from the northwestern rim, 
the southwestern rim and the eastern region of G347.3$-$0.5 could all
be adequately fit using a simple power law (with a photon index of 
$\approx$ 2). These fits were made based on $\it{ROSAT}$ and $\it{ASCA}$
observations (extending to approximately 10 keV), and we performed a fit 
to the $\it{RXTE}$ data using a power law with approximately the same 
index. We included a second component, a Gaussian with a line width 
$\sigma$ frozen to 0 keV, to fit the apparent emission feature
seen near 6.4 keV. We discuss this emission feature in more detail in
Section \ref{IronFeatureSubSection}. 
\par
\begin{figure}
\figurenum{3}
\plotone{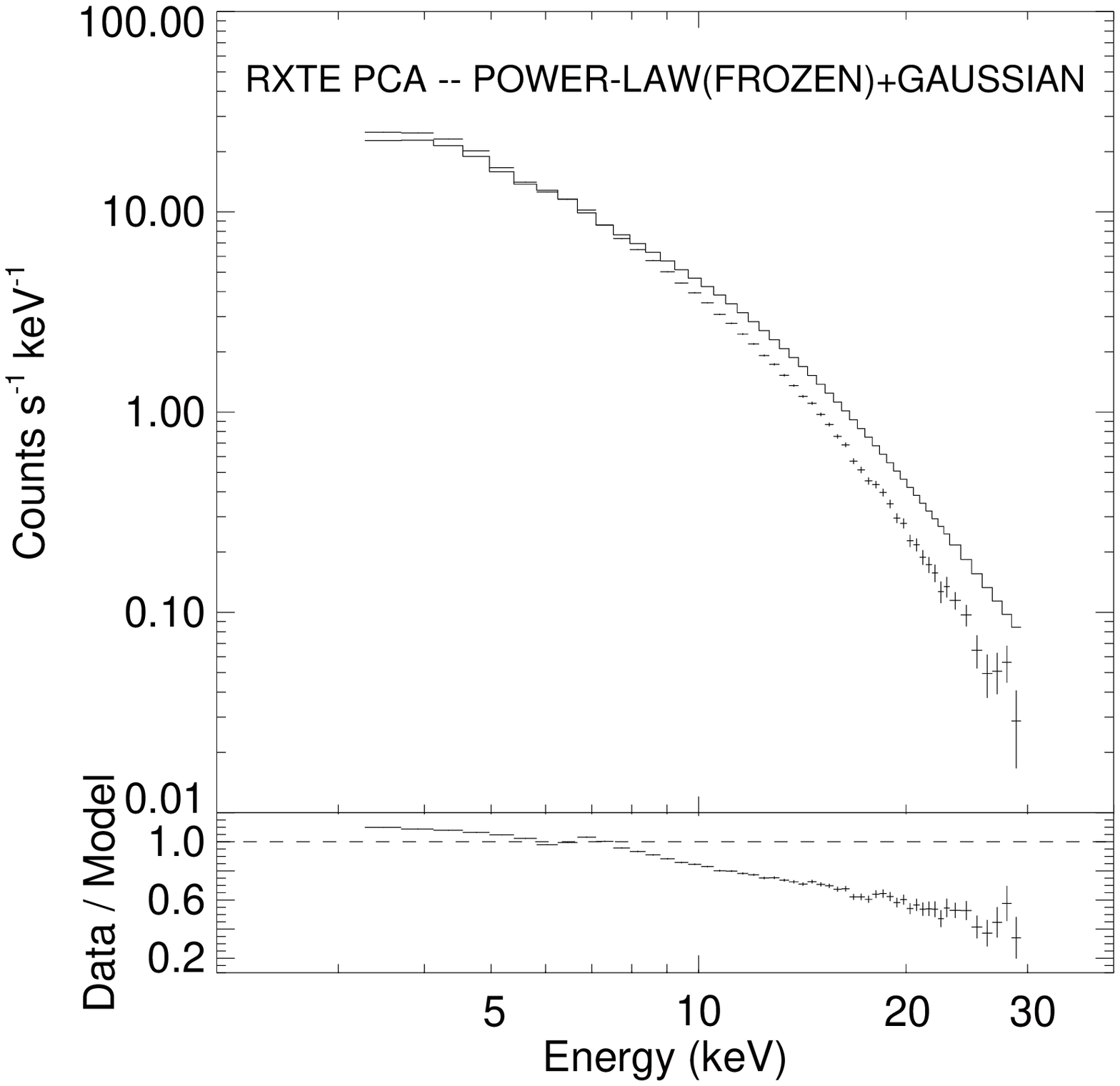}
\caption{Power Law (Frozen Index) + Gaussian fit to the {\it RXTE}~PCA
observations of G347.3$-$0.5.\label{fig3}}
\end{figure}
\begin{figure}
\figurenum{4}
\plotone{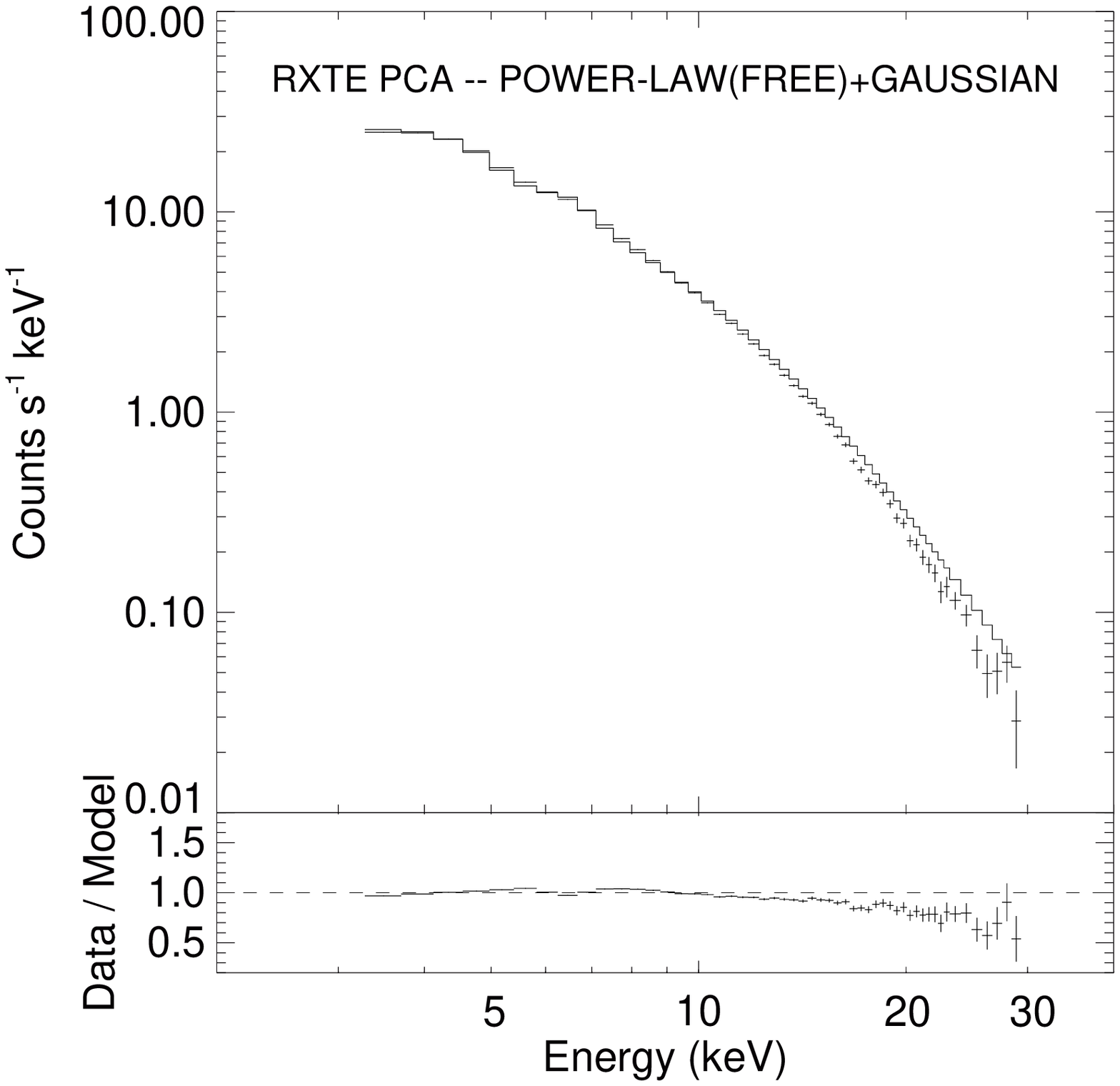}
\caption{Power Law (Free Index) + Gaussian fit to the {\it RXTE}~PCA 
observations of G347.3$-$0.5.\label{fig4}}
\end{figure}

Figure \ref{fig3} shows a POWER LAW+GAUSSIAN fit to the spectrum extracted 
from the $\it{RXTE}$ PCA observations, using an photon index of 2.33 to 
correspond to the weighted mean value of photon indices obtained in the 
fits made to the northwestern rim, southwestern rim and eastern region of 
G347.3$-$0.5 by \citet{S99}. Clearly, this model did not fit the data in 
this energy range, yielding a reduced $\chi$$^2$ of 482. Even when the 
photon index was a free parameter, an acceptable fit was still not obtained: 
in Figure \ref{fig4}, we plot the results of a POWER LAW+GAUSSIAN fit 
where we fit the photon index. In this Figure, the photon index is 2.61,
yielding a $\chi$$^2$ of 1808 for 49 degrees of freedom (a reduced
$\chi$$^2$ of 36.9). While this is certainly an improvement, this
model still does not fit the data. We conclude that the power law models 
employed by \citet{S99} over the energy ranges sampled by $\it{ROSAT}$ and
$\it{ASCA}$ cannot be extended into the $\it{RXTE}$ energy range. 
The energy distribution of the electrons that produce the observed
X-ray spectrum is therefore inconsistent with a simple power law. 
\par
$\it{SRESC}$: The second model used in our analysis was $\it{SRESC}$,
which describes a synchrotron spectrum from an electron distribution 
limited by particle escape above some energy. This model was used
by \citet{Dyer01} to sucessfully fit the X-ray spectrum of SN 1006,
and thorough descriptions of this model are presented by \citet{R96} and 
\citet{R98}. The $\it{SRESC}$ model describes electrons which are 
shock-acclerated in a Sedov blast wave encountering a constant-density 
medium containing a uniform magnetic field. This model also includes 
variations in electron acceleration efficiency with shock obliquity and 
post-shock radiative and adiabatic losses. 
\par
\begin{figure}
\figurenum{5}
\plotone{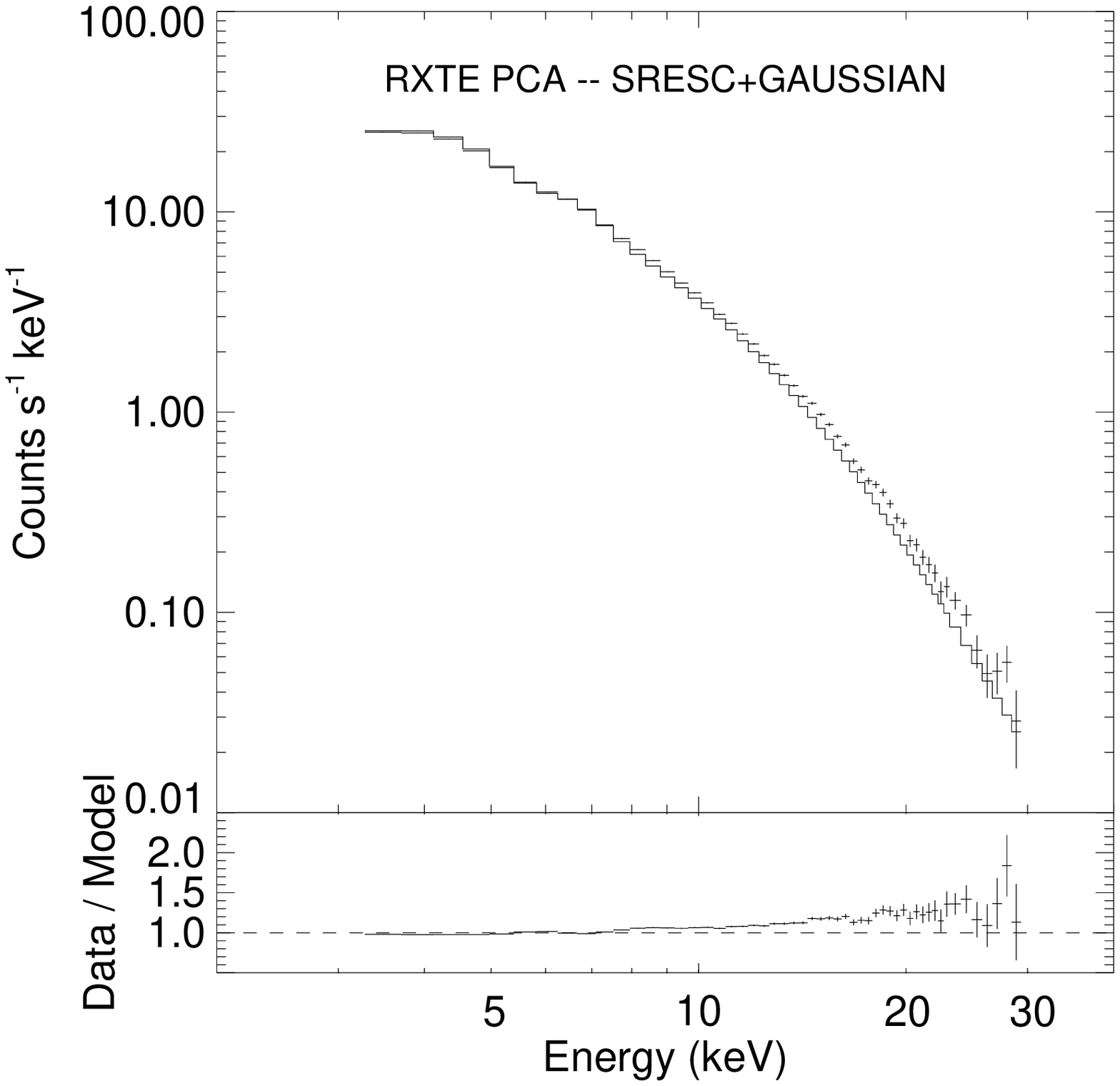}
\caption{SRESC + Gaussian fit to the {\it RXTE}~PCA observations
of G347.3$-$0.5.\label{fig5}}
\end{figure}

 
In Figure \ref{fig5}, we present a fit to the $\it{RXTE}$ spectrum using 
the $\it{SRESC}$ model combined with the $\it{GAUSSIAN}$ model. 
Similar to our attempts to fit the spectrum with a 
$\it{POWER~LAW}$+$\it{GAUSSIAN}$ model, we could not obtain
an acceptable fit to the $\it{RXTE}$ spectrum of G347.3$-$0.5 using this 
model. For the plotted model, the corresponding ratio of $\chi$$^2$
to degrees of freedom is 2451/48 = 51.07. We therfore conclude that
the $\it{SRESC}$ model, like the $\it{POWER~LAW}$ model, does not
properly describe the non-thermal X-ray emission seen from this SNR.
\par
$\it{SRCUT}$: The third model we considered is denoted as $\it{SRCUT}$:
it was described by \citet{R98}, \citet{RK99} and \citet{HR01}, 
with the latter two papers implementing the model in their study of the
maximum energies of electrons accelerated by samples of SNRs in the Galaxy 
and the Large Magellanic Cloud, respectively. The $\it{SRCUT}$ model
describes a synchrotron spectrum from an exponentially cut off power law 
distribution of electrons in a uniform magnetic field. The photon spectrum 
is itself a cut-off power law, rolling off more slowly than an exponential 
in photon
energies. Though this model is an oversimplification, it is more
realistic than a power law and it does give the maximally
curved physically plausible spectrum. This spectrum can in turn
be used to set limits on maximum accelerated electron energies
even in remnants whose X-rays are thermal. The $\it{SRCUT}$ model
assumes an electron energy spectrum $N$$_e$($E$) of the form
\begin{equation}
N_e (E) = K~E^{-\Gamma}~e^{-E/E_{cutoff}}
\end{equation}
where $K$ is a normalization constant derived from the observed
flux density of the region of the SNR at 1 GHz, $\Gamma$ is defined as
2$\alpha$+1 (where $\alpha$ is the radio spectral index) and 
$E$$_{cutoff}$ is the maximum energy of the accelerated cosmic-ray 
electrons. A crucial advantage of this model (as well as the 
$\it{SRESC}$~model) is 
that a resulting fit may be compared with two observable properties of an 
SNR, namely its flux density at 1 GHz and $\alpha$. Finally, one of
the fit parameters for the $\it{SRCUT}$~and $\it{SRESC}$~models is  
the cutoff frequency $\nu_{cutoff}$ of the synchrotron spectrum of the 
electrons. This quantity is defined as the frequency
at which the flux has dropped by a factor of 10 from a straight 
power law. We can express $\nu_{cutoff}$ in a way which is quantitatively 
consistent with previous work \citep{R98,RK99,HR01} as
\begin{equation}
\nu_{cutoff} \approx 0.48 \times 10^{16} \left( \frac{B_{\mu G}}{10~\mu G}
\right) \left( \frac{E_{cutoff}}{10~\mbox{TeV}} \right)^2
\quad \mbox{Hz,} \quad 
\label{nucutoffeqn}
\end{equation}
where $B_{\mu G}$ is the magnetic field strength of the SNR in
$\mu$G, assuming the electrons are moving perpendicular to the
magnetic field. By using the value 
for $\nu_{cutoff}$ returned by  $\it{SRCUT}$ as well as the normalization 
factor $K$, the synchrotron spectrum of the shock-accelerated electrons 
can be adequately described. Moreover, an estimate for the maximum
energy $E$$_{cutoff}$ for the shock-accelerated electrons can also
be derived.
\par
In Figure 6, we present a fit to the $\it{RXTE}$ spectrum using the 
$\it{SRCUT}$ model. Clearly, this model gives the best-fit to the
spectrum of G347.3$-$0.5 over this energy range: the corresponding
value for $\chi$$^2$/degrees of freedom is 207.98/48 = 4.33, which is
a considerable improvement (though still not statistically acceptable) 
over the fits obtained by both the
$\it{POWER~LAW}$ model and the $\it{SRESC}$ model. The normalization $K$ 
for the $\it{SRCUT}$ component of this fit indicates that the flux density 
at 1 GHz for the entire SNR is 6.9 $\pm$ 0.2 Jy (90\% confidence limits). 
This value is marginally consistent with the
weak radio emission seen from this SNR, known to be dominated by
a source of emission at the northwestern rim of G347.3$-$0.5 with
a corresponding flux density of 4 $\pm$ 1 Jy \citep{ESG01}. Moreover,
the broad field of view of the PCA certainly intercepts diffuse X-ray
flux from other sources besides G347.3$-$0.5, which leads to a 
corresponding
increase in the value for $K$ returned by the fit. We also note that by
using the best fit value for $\nu$$_{cutoff}$, Equation \ref{nucutoffeqn} 
yields $E$$_{cutoff}$ $\approx$ 68 TeV (assuming a magnetic 
field strength of 10 $\mu$G). This is in good agreement 
with the value of $E$$_{cutoff}$ obtained by \citet{ESG01} using
a similar value for $B$$_{\mu G}$ with different models. Based on our 
success in fitting the $\it{RXTE}$ spectrum of G347.3$-$0.5 with the
$\it{SRCUT}$ model, we conclude that the energy distribution of
the highest energy cosmic-ray electrons accelerated by this SNR is
consistent with an exponentially cut-off power law. 
\subsection{Joint Spectral Fitting of $\it{ROSAT}$ PSPC, $\it{ASCA}$ GIS
and $\it{RXTE}$ PCA Data\label{JointSpectralFitSubSection}}
We next applied the $\it{SRCUT}$ model to the X-ray spectrum
of G347.3$-$0.5 as observed by the $\it{ROSAT}$~PSPC and the $\it{ASCA}$~GIS
as well as the $\it{RXTE}$~PCA. We took advantage of the spatial
resolving
capabilities of the $\it{ROSAT}$~PSPC and the $\it{ASCA}$~GIS in order
to simultaneously fit spectra extracted from particular regions of the
SNR. We simultaneously fit eight different X-ray spectra:
the spectra for the northwestern rim, the southwestern rim, the
northeastern rim and the diffuse central emission (with the two interior
point sources omitted) as observed by the $\it{ROSAT}$~PSPC, the
spectra for the northwestern rim, the southwestern rim and the 
northeastern
rim as observed by the $\it{ASCA}$~GIS and finally the spectra for
the whole SNR as observed by the $\it{RXTE}$~PCA. 
\par
Two models (each a combination of a thermal component and a non-thermal  
component) were used for the joint spectral-fitting
process -- namely a $\it{RAYMOND}$+$\it{SRCUT}$~model and an 
$\it{EQUIL}$+$\it{SRCUT}$ model -- in order to fit both the
thermal and non-thermal components respectively in the individual
specta. In their study of the X-ray spectrum of G347.3$-$0.5 based
on $\it{ROSAT}$~PSPC and $\it{ASCA}$~GIS data, \citet{S99} found
no evidence for thermal emission from any portion of the SNR. 
We revisted this issue by testing whether the joint fit to the individual 
spectra would be improved by the inclusion of a thermal component. 
Both thermal models describe the emission spectrum from a hot diffuse gas, 
including line emission from several elements, that is in a state of
collisional ionization equilibrium. Both of 
these models have a common set of parameters: a plasma temperature 
$\it{kT}$, an abundance $Z$, a redshift $z$ and an emission measure
$\it{EM}$. For the purposes of the present work, we froze the values of 
$Z$ and $z$ to unity (i.e., solar abundance) and zero, respectively.
The emission measure $\it{EM}$ is defined as 
\begin{equation}
\mbox{EM} \mbox{(cm$^{-5}$)}= \frac{10^{-14}}{4 \pi d^2} \int
n_e n_H dV,
\label{EMEqn}
\end{equation} 
where $d$ is the distance to the source in centimeters
and $n$$_e$ and $n$$_H$ are the electron and hydrogen densities,
respectively, in units of cm$^{-3}$. We can evaluate this expression
as follows: we recall that we have assumed a distance of 6 kiloparsecs
to G347.3$-$0.5 and we note that the angular size of this SNR is 65$\arcmin$ 
$\times$ 55$\arcmin$ \citep{Green01}, therefore the corresponding linear
radius is 1.65 $\times$ 10$^{20}$ cm. Assuming a spherical volume for
the SNR with a filling factor of one-quarter (that is, assumed only 
one-quarter of the total volume of the SNR is filled with thermal 
particles that produce the observed emission), as well as uniform values
for $n$$_e$ and $n$$_H$ throughout the volume, then Equation 
\ref{EMEqn} yields 
\begin{equation}
\mbox{EM} \mbox{(cm$^{-5}$)} = \left( \frac{10^{-14}}{4 \pi d^2} \right)
\left( \frac{1}{4} \right) \left( \frac{4}{3} \pi r^3 \right) n_e n_H = 
\mbox{(10.92 cm)}~n_e n_H
\label{EMdEqn}
\end{equation}
We can estimate the ambient densities of hydrogen surrounding G347.3$-$0.5 
from the values obtained from the spectral fits for $\it{EM}$
(assuming $n$$_e$ = 1.2 $n$$_H$). We will continue this discussion when
we describe our detection of thermal X-ray emission from G347.3$-$0.5 in 
Section \ref{ThermalSubSection}. 
\par
We have modeled the background diffuse X-ray
emission from the GR as observed by the $\it{RXTE}$~PCA
with the same two component model used by \citet{VM98} mentioned 
earlier, namely a $\it{RAYMOND}$+$\it{POWER~LAW}$~model where the 
abundance 
$Z$ is frozen to solar values and the redshift $z$ is frozen to zero. 
We froze the values for $\it{kT}$, photon index and $N$$_H$~to be 2.9 keV, 
1.8 
and 1.8 $\times$ 10$^{22}$ cm$^{-2}$, respectively: these are the same 
best-fit values measured by \citet{VM98}. The normalizations for the 
$\it{RAYMOND}$~and $\it{POWER~LAW}$ components were left as free 
parameters. 
\par
In Figures \ref{fig7} through \ref{fig9}, we present data, fits
and residuals for the spectra extracted from the $\it{ROSAT}$~PSPC
regions (Figure \ref{fig7}), the $\it{ASCA}$~GIS regions (Figure
\ref{fig8}) and the whole SNR as observed by the $\it{RXTE}$~PCA 
(Figure \ref{fig9}) using the $\it{RAYMOND}$+$\it{SRCUT}$ model.
Similarly, in Figures \ref{fig10} through \ref{fig12}, we present
data, fits and residuals for the same set of spectra using the
$\it{EQUIL}$+$\it{SRCUT}$ model. In Tables \ref{SRCUTRAYFitTable}
and \ref{SRCUTEQUFitTable}, we present the fit parameters for
the depicted fits: the quality of the two fits is comparable for
the two models (a $\chi$$^2$/degrees of freedom of 808.25/471 = 1.72 
for the $\it{RAYMOND}$+$\it{SRCUT}$ model compared to a $\chi$$^2$/degrees 
of freedom of 843.76/472 = 1.79 for the $\it{EQUIL}$+$\it{SRCUT}$
model for 501 pulse height amplitude bins). We emphasize that, for both
two-component models, the fits successfully reproduce the observed
radio properties of G347.3$-$0.5, namely the two estimates for the
flux densities at 1 GHz for the northwestern rim are both 
consistent with the observed value of 4 $\pm$ 1 Jy \citep{ESG01}.
Also, the estimates for the flux densities at that frequency for
the other portions of the SNR are very modest, also consistent with
observations. In the next section we discuss these fit parameters in
more detail.




\section{Results\label{ResultsSection}}
\subsection{Cutoff Energies of Electrons Shock-Accelerated by 
G347.3$-$0.5\label{MaxEnergySubSection}} 
The electron energy $E$$_e$ associated with synchrotron photons
radiated at energy $E$$_X$ is given by the relation
\begin{equation}
E_e \approx \frac{80~\mbox{TeV}}{B_{\mu G}^{1/2}} \left( 
\frac{E_X}{1~\mbox{keV}} \right)^{1/2} 
\label{EEquation}
\end{equation}
We can use this equation and the values for $\nu$$_{cutoff}$ derived by 
our fits to estimate the maximum energy $E$$_{cutoff}$ of the electrons 
accelerated by G347.3$-$0.5. We took four values for $\nu$$_{cutoff}$ 
yielded by our fits to the emission from the northwestern rim
and the diffuse central emission using the 
$\it{SRCUT}$+$\it{RAYMOND}$ model and the $\it{SRCUT}$+$\it{EQUIL}$
model. Using Equation \ref{EEquation}, $E$$_X$ = $h$$\nu$$_{cutoff}$
yields an estimate for $E$$_{cutoff}$~(assuming a magnetic field of 10 
$\mu$G), and we list our calculated values for $E$$_{cutoff}$~in Table 
\ref{DerivedFitTable}. These values range from approximately
19 TeV to 25 TeV and are consistent with the values obtained
by \cite{ESG01}. We agree with the results and arguments of \citet{S99} 
and \cite{ESG01} that G347.3$-$0.5 is $\it{not}$~accelerating cosmic-ray
electrons to the ``knee" energy of 3000 TeV. In Section 
\ref{NWRimSection}, we describe a more rigorous attempt to estimate
the maximum energy of cosmic-ray electrons accelerated by this SNR,
where we simultaneously consider radio, X-ray and TeV emission from
the northwestern rim. We will employ a more sophisticated fitting
method where the ambient magnetic field strength is allowed to vary. 

\subsection{Thermal X-ray Emission from G347.3$-$0.5?
\label{ThermalSubSection}}

Previous X-ray observations of G347.3$-$0.5 by \citet{K97} and 
\citet{S99} found no evidence for thermal emission from any portion
of this SNR. \citet{S99} placed upper limits on the amount of thermal
emission as a function of $kT$ for both emission from the northwestern rim 
and the central diffuse emission of G347.3$-$0.5. By including a 
thermal component in fits to the X-ray emission from G347.3$-$0.5,
we re-examine the upper limits on thermal emission derived by 
\citet{S99}.
\par
In Table \ref{DerivedFitTable}, we present values for $E$$_{cutoff}$, 
$\Gamma$, $n$$_H$~and $n$$_e$ = 1.2 $n$$_H$, as calculated from our
model fit values for $\alpha$, $\it{EM}$ and Equation \ref{EMdEqn}.
We have considered only the fit values for the northwestern rim and the 
central diffuse
emission based on the results of both the $\it{SRCUT}$+$\it{RAYMOND}$ 
model and the $\it{SRCUT}$+$\it{EQUIL}$ model. For the northwestern
rim, we can only provide upper limits for $n$$_H$~and $n$$_e$. In
Figures \ref{fig13} and \ref{fig14}, we plot our values for $\it{EM}$ 
for the northwestern rim and the central diffuse 
emission, respectively, against the upper limits derived by \citet{S99}.
We find that for the northwestern rim, our values for $\it{EM}$ fall 
well below the upper limits of \citet{S99}, but
for the diffuse emission form the central region, our values for 
$\it{EM}$ more closely straddle the upper limits, and in the case of 
$\it{SRCUT}$+$\it{RAYMOND}$ our value exceeds the upper limit. 
This result indicates the first detection of thermal X-ray
emission from G347.3$-$0.5: we speculate that earlier work may have
missed this thermal emission because earlier efforts had concentrated
on the non-thermal properties of this SNR (particularly the bright
rims of this SNR) and only considered a narrower energy range of the
spectrum of G347.3$-$0.5. In addition, the rather large angular size of
this SNR makes it difficult to completely sample all of its X-ray
properties. A more thorough understanding
of the rather modest thermal emission from this SNR will require
additional observations and modeling.

\subsection{The Detection of Iron Emission -- Galactic Ridge 
Background?\label{IronFeatureSubSection}}

We now comment on a modest spectral feature seen near 6.4 keV
in the $\it{RXTE}$~PCA spectrum of G347.3$-$0.5 and at a similar 
energy in the spectra from the two background observations also made with
the $\it{RXTE}$~PCA. This emission feature may be associated with iron,
which is known to produce a broad set of emission features between the
energies of 6 keV and 7 keV. One possibility to consider is that
the feature is produced by thermal emission from G347.3$-$0.5: we note that 
\citet{Bykov02} has recently argued that fast moving isolated fragments of 
SN ejecta composed of heavy elements (such as iron) should produce 
K$\alpha$ fluoresence emission lines at X-ray energies, and that the
iron K$\alpha$ fluoresence emission line energy is known to be 
approximately 6.4 keV. A second possibility is that the feature 
originates from diffuse emission associated with the GR, which is seen in 
projection beyond G347.3$-$0.5. The large field of view of the 
$\it{RXTE}$~PCA (approximately 1$^{\circ}$ FWHM) and the lack of imaging 
capabilities of this instrument make it difficult to clearly associate 
this feature with either G347.3$-$0.5 or the GR. Recently, \citet{Tanaka02} 
presented an analysis of spectral features -- including the iron K$\alpha$
fluoresence emission line -- associated with different portions of the GR,
as observed by the $\it{ASCA}$~SIS. 
\par
In Table \ref{RXTEPCATable}, we present the line energies and 
normalizations (along with the 90\% confidence limits for these values) 
for the emission feature as seen in the $\it{RXTE}$~PCA spectra of 
G347.3$-$0.5 and the two background pointings. We also performed a 
simultaneous fit of the emission feature as seen in the three spectra, 
where the line energies and the normalizations for the feature as seen
in each spectra were fixed to be the same. We also 
present the best-fit values for the line energy and the normalization 
for the joint fit as well. Notice that in each case, the line energies and
the normalizations are all approximately the same, within the confidence
limits. In order to check for any instrumental effects that may be 
corrupting the data, we obtained archived data from an $\it{RXTE}$~PCA 
observation of Cas A made on 1999 August 5 (OBSID 40806$-$01$-$04$-$00):
we selected this particular observation of Cas A for analysis because it 
took place nearly contemporaneously to our G347.3$-$0.5 observation
(less than two months later). Using the same data
reduction techniques described in Section \ref{RXTESubSection}, we
generated a source spectrum of Cas A and measured both the line
energy and the normalization of the iron K$\alpha$ emission line
seen in the spectrum of Cas A. We measured a line energy 
of 6.66$^{+0.01}_{-0.02}$ keV, which is in good agreement with the 
energy of this emission line in the $\it{ASCA}$~SIS spectra of
Cas A as observed by \citet{Holt94}. Thus, we conclude that there
are no instrumental effects that are corrupting our analysis of the
emission feature near 6.4 keV. 
\par
We therefore hypothesize that the spectral feature seen in the 
$\it{RXTE}$~PCA is not associated with G347.3$-$0.5 for two major reasons.
First, this feature was not detected by either \citet{S99} or by us during 
the analysis of $\it{ASCA}$~GIS observations of this SNR which sampled the 
energy range which includes this spectral feature. Second, by inspection 
of Figure \ref{fig2} and Table \ref{RXTEPCATable}, it is clear that 
the emission feature is seen in both the on-source $\it{RXTE}$~PCA 
pointing and the two background pointings, and the line energies and 
normalizations of the emission features for all three pointings are similar. 
If this feature was associated with G347.3$-$0.5 rather than the GR,
we would expect that the line energy and/or the normalization in the source
region would significantly differ from the values seen for the feature in 
the two background pointings. In the study of spectral features observed 
from diffuse X-ray emission from the GR, \citet{Tanaka02}
found that only diffuse emission from within 1$^{\circ}$ of the Galactic 
Center exhibited a feature near 6.4 keV, while the diffuse line spectrum 
from two other regions -- centered at longitudes of 10$^{\circ}$ and 
28.5$^{\circ}$, respectively -- were dominated instead by a feature near 
6.7 keV. Because neither of these two regions include the positions of 
G347.3$-$0.5 or the two background regions, however, we argue that the 
results from \citet{Tanaka02} are not applicable to our study presented 
here. We therefore conclude that the spectral feature is not associated 
with G347.3$-$0.5 and instead originates in the diffuse X-ray emission 
from the GR seen in projection beyond G347.3$-$0.5.

\subsection{X-ray Pulsations from the Radio Pulsar PSR J1713$-$3949 
(Possibly Associated with 1WGA J1713.4$-$3949 and  
G347.3$-$0.5)?\label{PulsarSearchSubSection}}

Recent radio observations of 1WGA J1713.4$-$3949 (the X-ray source
centrally located inside G347.3$-$0.5) by \citet{Crawford02} discovered
a new pulsar (PSR J1713$-$3949) with a period of 392 ms within 
7 arcminutes (the discovery beam radius) of the X-ray source. 
\citet{Crawford02} derived a distance of 5.0$\pm$0.2 kiloparsecs
to this source, which is consistent with the distance estimate to
G347.3$-$0.5 of \citet{S99}. The position of 1WGA J1713.4$-$3949 near the
center of G347.3$-$0.5 suggests that the X-ray source (as well as 
the radio pulsar) and the SNR are associated. In addition, the position
of G347.3$-$0.5 itself near a molecular cloud also suggests that the
SNR was produced by a massive star in a core collapse SNe: these types
of SNe are expected to produce neutron stars. To further investigate the
nature of 1WGA J1713.4$-$3949, we performed a timing analysis of our
$\it{RXTE}$~PCA data to search for pulsations from this source.
\par
In Figure \ref{fig15}, we present a power spectrum of our $\it{RXTE}$~PCA 
data over the frequency range of 2.5 through 2.6 Hz with a dashed
vertical line indicating a period of 392 ms (2.55 Hz). We find no
evidence for pulsations in our spectrum at any frequency, including
2.55 Hz. Our failure to detect pulsations from this central source
may be explained by several factors, such as significant contaminating 
diffuse background emission from the GR, cosmic ray background and 
finally emission from the SNR dominating over emission
from a pulsar over the energy range sampled by our $\it{RXTE}$~PCA
observation. Pointed high angular resolution X-ray observations 
would be much more sensitive to pulsed X-ray emission from this source.

\section{The Broadband Energy Spectrum of the Northwestern Rim
of G347.3$-$0.5\label{NWRimSection}}

In Figure \ref{fig16}, we present a broadband energy spectrum of the 
northwestern rim of G347.3$-$0.5 ranging from radio through $\gamma$-ray
energies. We have constructed this diagram using published values for
the 1 GHz radio observation of G347.3$-$0.5 by the Australian 
Telescope Compact Array (ATCA) as presented by \citet{ESG01}, the
$\it{ROSAT}$~PSPC and $\it{ASCA}$~GIS data described previously in
this paper and data from the CANGAROO observation at TeV energies of the 
spectrum of this rim as published by \citet{Enomoto02}. Using the
$\it{ISIS}$~software package\footnote{See http://space.mit.edu/ASC/ISIC/
for more information about $\it{ISIS}$.} \citep{HD00}, we obtained a fit
to the energy spectrum using a model that includes multiple
high energy processes associated with SNRs, as indicated by separate
curves. These curves indicate the emission profile expected by a 
shock-accelerated population of electrons from two different 
processes: synchrotron emission (as indicated by the curve labeled ``S"), 
and inverse-Compton emission from cosmic microwave background photons that 
have been upscattered to TeV energies by energetic cosmic-ray electrons 
(the curve labeled ``IC"). These emission models are consistent with the
models employed by \citet{Sturner97}, and a detailed discussion of the
high energy emission from non-linear shock acceleration processes 
associated with SNRs is given by \citet{Baring99}. From our fit to the 
$\it{ASCA}$~GIS data and the TeV data, we have derived estimates
for $E$$_{cutoff}$ for the accelerated cosmic-ray electrons and 
magnetic field strength $B$. For the depicted fit, the ratio of $\chi$$^2$ 
to the degrees of freedom is 191.78/113 = 1.70. Based on this fit, 
we estimate a value of only 8.8$_{-3.4}^{+4.1}$ TeV for 
$E$$_{cutoff}$ of the accelerated cosmic-ray electrons and a corresponding 
magnetic field strength of $B$ = 150$_{-80}^{+250}$ $\mu$G. This value for
$E$$_{cutoff}$ is sharply lower (but still consistent) with the 
upper limit derived by \citet{ESG01}, while the corresponding value
for $B$ is considerably larger than the value of 10$\mu$G assumed by 
\citet{S99}. There is also a modest amount of overlap between our range of 
values for $B$ and the the range suggested by \citet{Enomoto02} (10 $\mu$G --
100 $\mu$G). However, the physical plausibility of this fit may be
questioned based on the value derived for the ratio of the volumes for TeV 
emission and X-ray emission from the northwestern rim -- that is, 
$V$$_{TeV}$/$V$$_{X-ray}$. In contrast to \citet{Enomoto02}, who fixed
this ratio to equal unity, we allowed this ratio to vary and obtained
a value of $V$$_{TeV}$/$V$$_{X-ray}$ $\approx$ 1000, with a lower limit
of 360, assuming that the power law index is 2.1 (derived from our 
from our model fits where $\alpha$ = 0.55) and that there is no curvature
to the electron spectrum. This ratio is too high to be physically
acceptable, and suggests that the TeV emission from the northwestern
rim of G347.3$-$0.5 is not produced by inverse Compton scattering of 
cosmic microwave background photons off high energy electrons accelerated 
along this rim of the SNR. 
\par
Nonetheless, while this TeV emission may not be produced by inverse
Compton scattering, we argue that neutral pion decay also cannot
produce this emission. We point out that in arguing that the TeV emission
is produced by neutral pion decay, \citet{Enomoto02} assumed an 
extremely large ambient density ($n$ $\approx$ 100 cm$^{-3}$ if a distance 
of 6 kpc to G347.3$-$0.5 is assumed) for the northwestern rim of the SNR. 
Our fits to the thermal component of the X-ray emission from the northwestern 
rim and the central diffuse emission yield a dramatically lower ambient 
density ($n$ $\approx$ 0.05-0.07 cm$^{-3}$) and make the conclusions 
of \citet{Enomoto02} seem untenable. \citet{S99} also concluded that
G347.3$-$0.5 is expanding into an ambient interstellar medium with
an extremely low density, and suggested that the SNR is expanding
into the evacuated wind-blown cavity created by the stellar progenitor:
we agree with this finding, and suggest that while G347.3$-$0.5 may have
formed in the vicinity of dense molecular clouds, it has not yet begun to 
interact with these clouds. We also note that the observed TeV emission
is observed to be centered on the northwestern rim of the SNR and 
{\it not}~on the adjacent cloud, as assumed by \citet{Enomoto02}. Finally,
\citet{RP02} noted that if the 
flux of the GeV emission from the $\gamma$-ray source 3EG J1714$-$3857 can 
be used to constrain the 
neutral pion decay spectrum fit presented by \citet{Enomoto02}: if this
source is associated with the northwestern rim of G347.3$-$0.5, its flux 
must be taken into account in all spectral fits, while if the $\gamma$-ray 
source is not associated with this rim, then the GeV emission may be 
treated instead as upper limits on flux along this energy range. 
\citet{RP02} showed that the flux from the neutral pion decay spectrum
fit exceeded the observed GeV emission by a factor of three, therefore
strongly arguing against neutral pion decay as the process responsible
for the TeV emission. For these reasons, we argue that the TeV emission
observed from the northwestern rim of G347.3$-$0.5 is not produced by
neutral pion decay. We also comment that if the ambient density is as
low as inferred by our estimates, the process of non-thermal 
bremsstrahlung emission -- as well as neutral pion decay -- will be
significantly suppressed. A non-thermal bremsstrahlung origin for the
TeV emission may also be ruled out because a spectrum generated by
this process would -- like the spectrum produced by neutral pion 
decay -- exceed the observed GeV emission. In conclusion, we state that 
none of the
known high energy processes associated with SNRs -- inverse Compton
scattering, neutral pion decay and non-thermal bremsstrahlung -- 
can adequately fit the observed TeV emission from the northwestern
rim. From the work of \citet{E00}, we suspect that all models
of high energy emission from SNRs over broad energy ranges must take
into account some curvature in the energy spectrum of accelerated 
electrons rather than assuming simple power laws. In such a case, one
of these emission processes may be able to adequately fit the observed
TeV emission from the northwestern rim of G347.3$-$0.5 -- and perhaps
the rims of other SNRs detected at such high energies -- after all. 
We will discuss such
curvature in the energy spectra of SNRs in more detail in a future work.
\par
Lastly, we consider the putative association between G347.3$-$0.5
and the X-ray source AX J1714.1$-$3912, a hard spectrum source
coincident with a molecular cloud located along the northeastern
rim of G347.3$-$0.5. \citet{Uchiyama03} presented an analysis of the
spectral properties of this source using data from ASCA 
$\it{GIS}$~observations. The X-ray spectrum of this source was best
fit by a power law with a rather flat photon index of 
$\Gamma$=1.0$^{+0.4}_{-0.3}$, prompting \citet{Uchiyama03} to argue 
that such a spectrum was best interpreted as non-thermal bremsstrahlung
emission from particles accelerated at the shock interface between
the northeastern rim of G347.3$-$0.5 and the molecular cloud. Despite
the spatial proximity between G347.3$-$0.5 and the molecular cloud
(and, by extension, AX J1714.1$-$3912), the evidence for the association 
between these sources is not compelling because, based on simulations 
of the interactions between SNRs and interstellar clouds \citep{JJ99},
we expect the amount of radio emission from both the SNR and the
cloud to dramatically increase as a result of the interaction.
\citet{JJ99} also predict that if the SNR has overtaken the cloud, 
the radio emission from the SNR would ``wrap around" the cloud, and
such a morphology should be readily apparent. 
Radio images of G347.3$-$0.5 and its surrounding environment have been
made with the Molonglo Observatory Synthesis Telescope (MOST)
\citep{R94} and presented by \citet{S99} (at the frequency of 843 MHz)
and \citet{ESG01} (at the frequency of 1.4 GHz). Inspection of these
images reveals little (if any) radio emission at either the position
of AX J1714.1$-$3912 or the northeastern rim of G347.3$-$0.5, and
no evidence for a ``wrap around" morphology in the radio emission.
The failure to detect radio emission from the hard X-ray source or
the northeastern rim of the SNR and a ``wrap around" morphology to the
radio emission argues against an interaction taking place at that site. 
Another reason to doubt that an interaction is taking place between 
G347.3$-$0.5 and the molecular cloud is related to the ratio of 
emission from rotational transitions of CO molecules,  
CO($J$=2$\rightarrow$1)/CO($J$=1$\rightarrow$0), as measured at
the site of the cloud. An enhanced value for this ratio (such as
a value of 1.2 or greater) is thought to indicate that an interaction
is taking place between the cloud and a shock, such as the one 
associated with an SNR \citep{Seta98}. Both \citet{S99} and \citet{Butt01}
found an enhanced value for this ratio for the cloud along the 
northeastern rim of the SNR, but we suggest that in this particular
case the enhanced CO($J$=2$\rightarrow$1)/CO($J$=1$\rightarrow$0) ratio
and the low ambient density of the SNR cannot be reconciled, and that
(as mentioned before) G347.3$-$0.5 is not currently interacting with this 
molecular cloud. \citet{Butt01} measured a value for the flux ratio 
CO($J$=2$\rightarrow$1)/CO($J$=1$\rightarrow$0) of 2.4$\pm$0.9:
according to \citet{DR84}, such a flux ratio corresponds to a shock 
velocity of less than 5 km/sec in the cloud. The shock speed in a cloud 
is reduced by the square root of the ratio of the ambient densities.
Based on CO observations \citep{Bronfman89, Butt01}, the mean density
of the cloud has been estimated to be approximately 500 cm$^{-3}$. 
From our spectral fits, we have inferred an average ambient density  
G347.3$-$0.5 to be $\approx$ 0.06 cm$^{-3}$: from these estimates for
the mean density of the cloud and ambient density around the SNR, we
conclude that the shock velocity outside the cloud is less than 500 
km/sec. For comparison, if we assume that the electrons and protons
in G347.3$-$0.5 are in thermal equilibrium at $\it{kT}$ = 1.5 keV
(from our fits to the diffuse thermal emission from the SNR), then 
the corresponding shock velocity is approximately 1100 km/sec. Since
the proton temperature is most likely much larger than the electron
temperature in the case of young SNRs \citep{Hwang02}, like 
G347.3$-$0.5, this value for the 
shock velocity is actually a lower limit, and therefore the shock velocity
exceeds 1100 km/sec, which is in sharp disagreement with the shock 
velocity of $<$ 500 km/sec derived earlier. \citet{Uchiyama03} noted that
the large X-ray luminosity of AX J1714.1$-$3912 (1.7 $\times$ 10$^{35}$
ergs/sec at the assumed distance of 6 kpc) made G347.3$-$0.5 barely
able to provide enough kinetic energy (in the form of nonthermal 
particles) to power this source, unless G347.3$-$0.5 is much closer
than 6 kpc or the SN explosion which created this SNR had more than
10$^{51}$ ergs of energy. We suggest that G347.3$-$0.5 is located
in the vicinity of the clouds, but has not yet started to interact with 
any of the clouds. More observations and analysis are required to 
determine the true nature of AX J1714.1$-$3912. 

\section{Comparison with Other Shell-Type SNRs with Non-Thermal
X-ray Emission: SN 1006 and G266.2-1.2\label{ComparisonSection}}

In Table \ref{ShellSNRsTable}, we compare the gross properties of   
G347.3$-$0.5 with two other dynamically-young shell-type SNRs which possess
significant non-thermal components to their X-ray emission, SN 1006  
\citep{Dyer01, A01} and G266.2$-$1.2 \citep{Slane01}. Inspection of this
Table \ref{ShellSNRsTable} reveals that all three SNRs are X-ray luminous,
radio faint and are expanding into regions of the interstellar medium with
particularly low ambient density. Both core-collapse and Type Ia SNe
appear to produce SNRs with non-thermal X-ray emission: SN 1006 is the  
archetype of a SNR produced by a Type Ia SN, while G266.2$-$1.2 and
G347.3$-$0.5 are thought to be produced by Type II SNe. The X-ray
morphologies of these three SNRs also contrast as well: for SN 1006,
the morphology is bilateral with two X-ray bright rims, visible in
$\it{ASCA}$~images of this SNR \citep{Dyer01}, while the morphologies
of both G266.2$-$1.2 and G347.3$-$0.5 feature multiple bright rims:  
$\it{ROSAT}$~and~$\it{ASCA}$ images of these two SNRs reveal
luminous northwestern, southwestern and northeastern rims for both
SNRs, while G266.2$-$1.2 also has a luminous southern rim as well
\citep{S99, Slane01}. The X-ray synchrotron flux depends on the 
exponential cut-off energy, the strength of the magnetic field, and the 
density of nonthermal electrons. In the case of SN 1006, the morphology 
may be due to an enhancement (via compression) in the magnetic field along 
the two bright rims relative to the magnetic field strength elsewhere 
along the rim. The morphology of SN 1006 -- with two symmetric luminous
X-ray rims -- was reproduced remarkably well by the work of 
\citet{R98}, who described models of synchrotron X-rays from
shell supernova remnants. In the cases of G347.3$-$0.5 and G266.2$-$1.2, the
cause of the morphology is not clear.  The bright rims may be
associated with regions where the density of nonthermal electrons is
relatively high. The common property of low ambient density for
these SNRs does not imply an exclusive requirement for non-thermal X-ray
emission: for example, another SNR, Cas A, possesses a strong non-thermal
X-ray component to its emission \citep{Vink03} -- though not a component
which dominates the X-ray spectrum of the SNR -- and is expanding into a
particularly dense ambient medium. Nonetheless, it appears that a low
ambient density -- independent of the type of SN progenitor -- is
conducive to the X-ray spectrum of the SNR being dominated by non-thermal
X-ray emission. Deep X-ray observations and detailed analysis of other
SNRs located in regions of low ambient density are required to determine
the degree to which low ambient density is necessary for producing this 
type of X-ray emission: we will explore this subject
in more detail using a sample of such SNRs in a future work.

\section{Conclusions\label{ConclusionsSection}}

The results and conclusions of this work may be summarized as
follows:
\par
1) We present a spatially-resolved X-ray spectral analysis of
both the three bright rims and the diffuse central emission of 
the Galactic SNR G347.3$-$0.5. This analysis involves data from
observations made of this source using instruments aboard three
different X-ray satellites (namely the $\it{ROSAT}$~PSPC, the
$\it{ASCA}$~GIS and the $\it{RXTE}$~PCA) spanning the approximate
energy range of 0.5 -- 30 keV.
\par
2) We have successfully fit the spatially-resolved X-ray spectra
of G347.3$-$0.5 using the $\it{SRCUT}$ model, describing an exponential
cut-off in the population of the highest energy electrons accelerated
by this SNR. In contrast, neither the $\it{SRESC}$ model nor a simple
power law model yielded acceptable fits to the data. 
We find that the power law fits obtained
by \citet{S99} to fit the spectrum of G347.3$-$0.5 over the energy 
range of 0.5-8 keV do not adequately fit the spectrum of this SNR
at higher energies. We have also detected for the first time a thermal 
component to the X-ray emission from this SNR: this component appears
to be more closely associated with central diffuse emission from the
SNR than from the X-ray luminous rims. We estimate the ambient density 
surrounding the SNR to be $\approx$0.05-0.07 cm$^{-3}$: this value is 
consistent with the range of values obtained by \citet{S99}. 
\par
3) A weak emission feature is seen near 6.4 keV in our $\it{RXTE}$~PCA
spectra of G347.3$-$0.5 and our two background pointings which sample the
GR. Because the strength and the line energy of this feature are 
approximately the same in all three pointings, we argue that the
feature is associated with diffuse background emission
from the GR rather than from G347.3$-$0.5 itself. 
\par
4) We have analyzed our $\it{RXTE}$~PCA data to search for X-ray pulsations
from the radio pulsar PSR J1713$-$3949, which lies near the center of
G347.3$-$0.5 and may be associated with the SNR. We cannot confirm the
presence of X-ray pulsations from this source, though our data is dominated 
by both cosmic ray background and X-ray emission from the SNR and the GR, 
making the detection of pulsed signal from the pulsar difficult. 
\par
5) Based on our two component (thermal and non-thermal) fits to the
X-ray emission from the northwestern rim and the central diffuse
emission from G347.3$-$0.5 and assuming a magnetic field strength of $B$ = 
10 $\mu$G, we estimate the maximum energy $E$$_{cutoff}$ of accelerated
cosmic ray electrons to be 19-25 TeV, consistent with previous 
analyses. Fitting the broadband (radio to $\gamma$-ray) energy spectrum 
of G347.3$-$0.5 with a synchrotron-inverse Compton scattering model yields 
values of 8.8$_{-3.4}^{+4.1}$ TeV for the maximum energy $E$$_{cutoff}$ of 
the accelerated cosmic-ray electrons and a magnetic field of
$B$ = 150$_{-80}^{+250}$ $\mu$G. Our value for $E$$_{cutoff}$ is
sharply lower (but still consistent) with upper limits derived by
previous studies, while our value for $B$ is significantly higher
than previously assumed. However, the ratio of volumes of TeV emission
to X-ray emission derived by this fit ($V$$_{TeV}$/$V$$_{X-ray}$
$\approx$ 1000, with a lower limit of 360) is too large to be physically
reasonable. While it appears that inverse Compton scattering cannot
adequately fit the TeV emission from the northwestern rim of G347.3$-$0.5, 
the other two high energy processes associated with SNRs -- neutral
pion decay, as argued by \citet{Enomoto02}, and non-thermal bremsstrahlung
-- can also be ruled out as well. This situation may change with the
application of more sophisticated models which take into account
curvature of the energy spectrum of electrons. 
\par
6) We have considered the gross properties of G347.3$-$0.5 and two
other SNRs known to feature X-ray spectra dominated by non-thermal
emission, SN 1006 and G266.2$-$1.2. We find that all three of these
SNRs are dynamically young, X-ray luminous but radio weak SNRs
expanding into regions of low ambient density. This suggests that
low ambient density may play a key role in dictating that the X-ray
spectrum from a SNR is dominated by non-thermal emission, though a 
more detailed study of the X-ray spectrum of other SNRs is required 
in order to test this hypothesis. 




\acknowledgments

We acknowledge useful discussions with Yousaf Butt, Kristy Dyer, 
Bryan Gaensler, Jasmina Lazendic, Joshua Migliazzo, Steven Reynolds,
Mallory Roberts, Patrick Slane and Diego Torres. We thank the 
anonymous referee for useful comments which have helped to enhance
the quality of this paper. This research has made 
use of data obtained from the High Energy Astrophysics Science Archive 
Research Center (HEASARC), provided by NASA's Goddard Space Flight Center. 
This research has also made use of NASA's Astrophysics Data System 
Abstract Service. For the data analysis, we have made use of the 
{\it ISIS} software package and the {\it LHEASOFT} package, which is
provided and maintained as a service of the Laboratory for High Energy 
Astrophysics (LHEA) at NASA/GSFC and the High Energy Astrophysics 
Division of the Smithsonian Astrophysical Observatory (SAO). T.G.P. 
acknowledges support from NASA LTSA grant NAG5-9237.

\clearpage

\clearpage

\clearpage

\clearpage

\clearpage

\clearpage
\begin{figure}
\figurenum{6}
\plotone{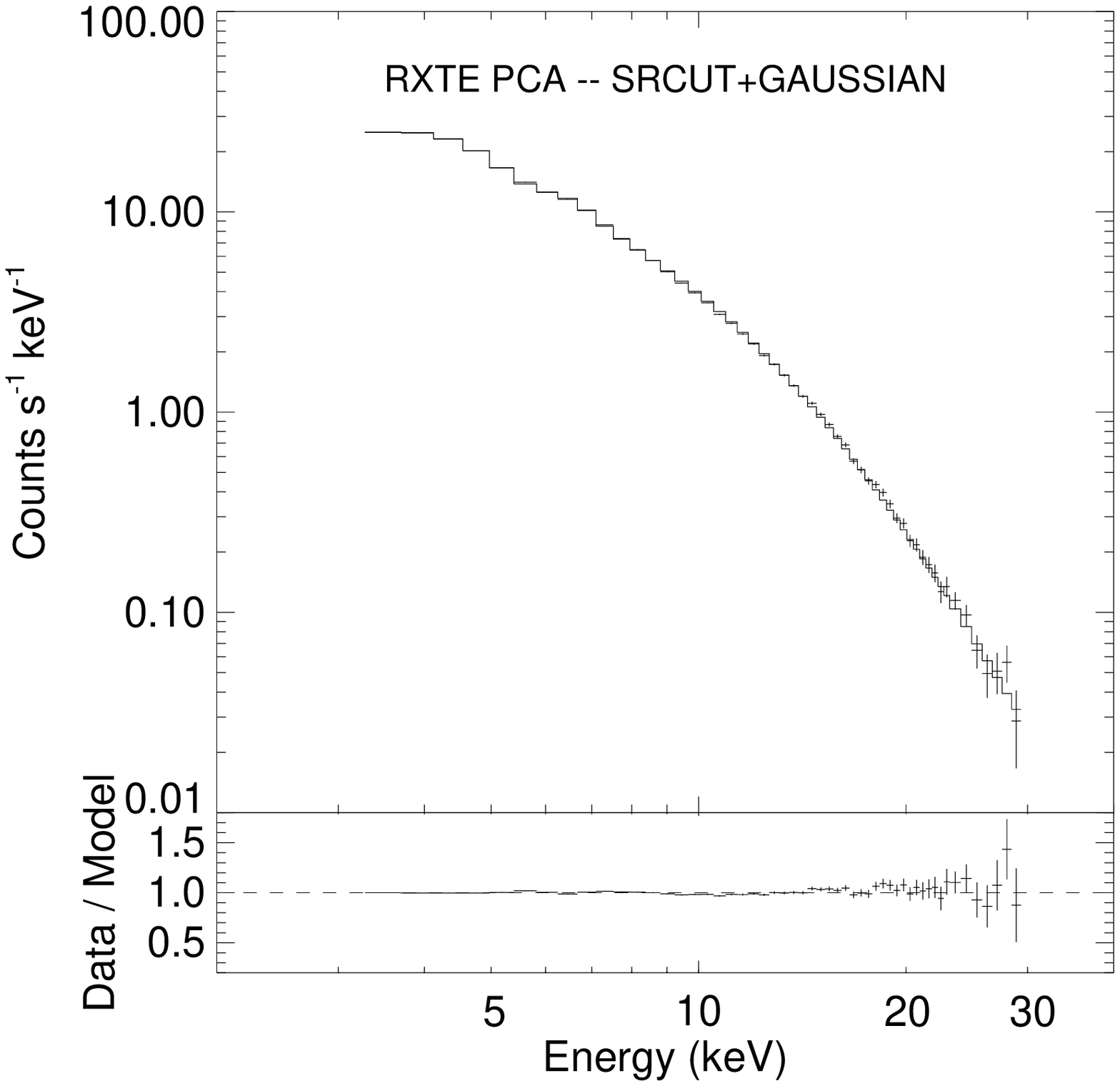}
\caption{SRCUT + Gaussian fit to the {\it RXTE}~PCA observations
of G347.3$-$0.5.\label{fig6}} 
\end{figure}

\clearpage
\begin{figure}
\figurenum{7}
\epsscale{0.80}
\plotone{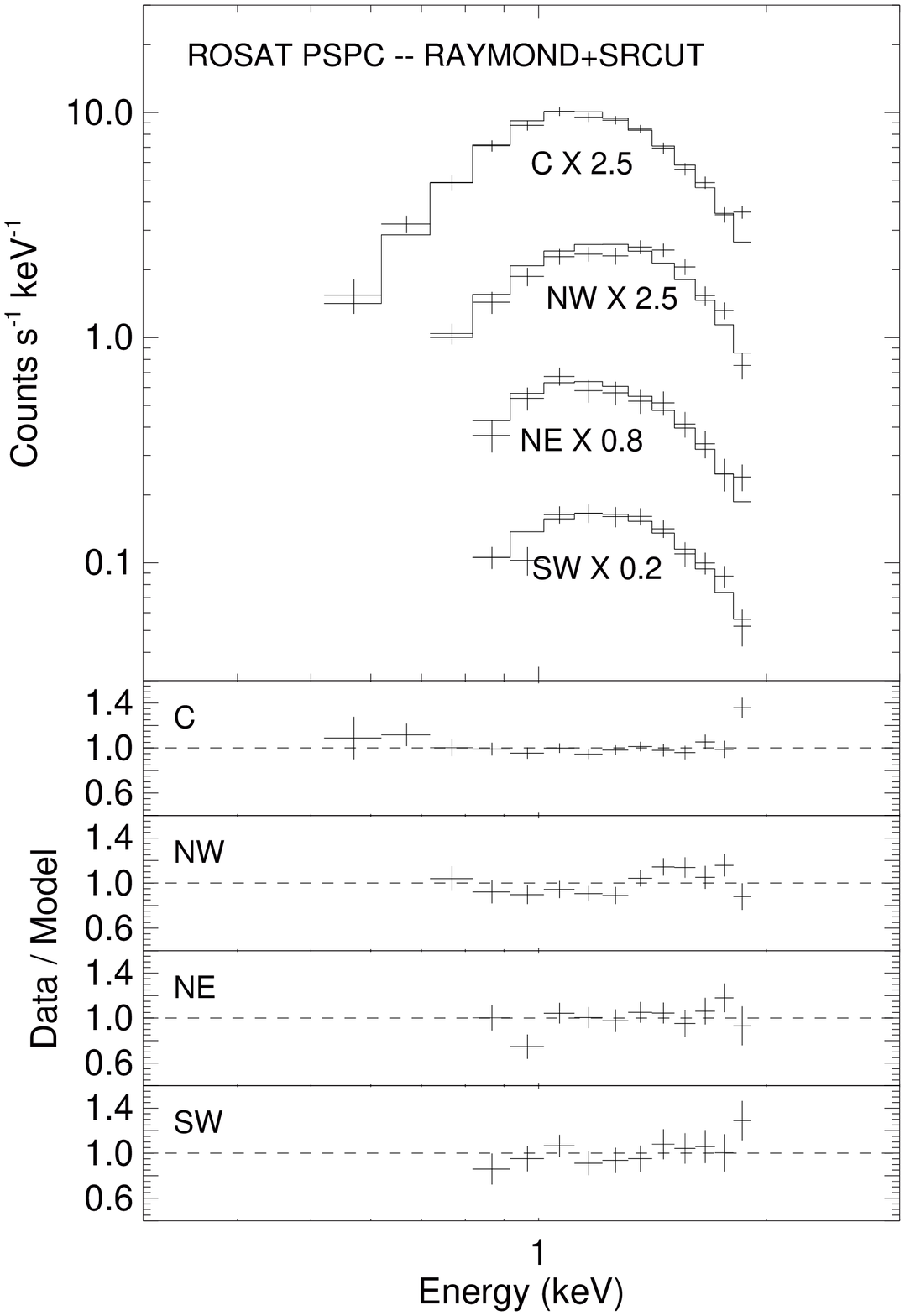}
\epsscale{1.00}
\caption{Spectra of the center (C), the northwestern rim (NW), the
northeastern rim (NE) and the southwestern rim (SW) of G347.3$-$0.5 
as observed by the {\it ROSAT}~PSPC and fit using the 
{\it RAYMOND}+{\it SRCUT} model.\label{fig7}} 
\end{figure}

\clearpage
\begin{figure}
\figurenum{8}
\plotone{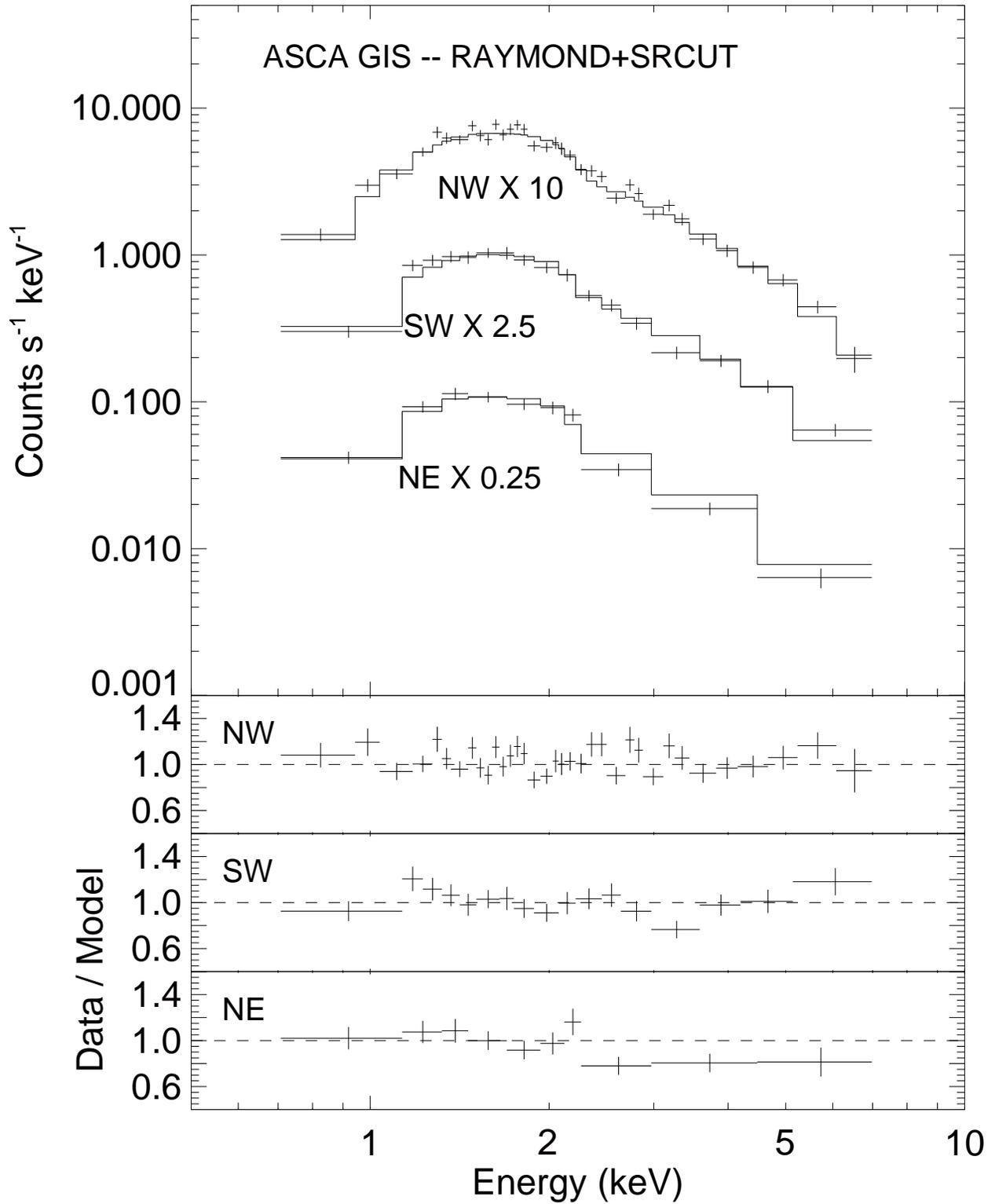}
\caption{Spectra of the northwestern rim (NW), the southwestern rim
(SW) and the northeastern rim (NE) of G347.3$-$0.5 as observed by the 
{\it ASCA}~GIS and fit using the {\it RAYMOND}+{\it SRCUT} 
model.\label{fig8}}
\end{figure}

\clearpage
\begin{figure}
\figurenum{9}
\plotone{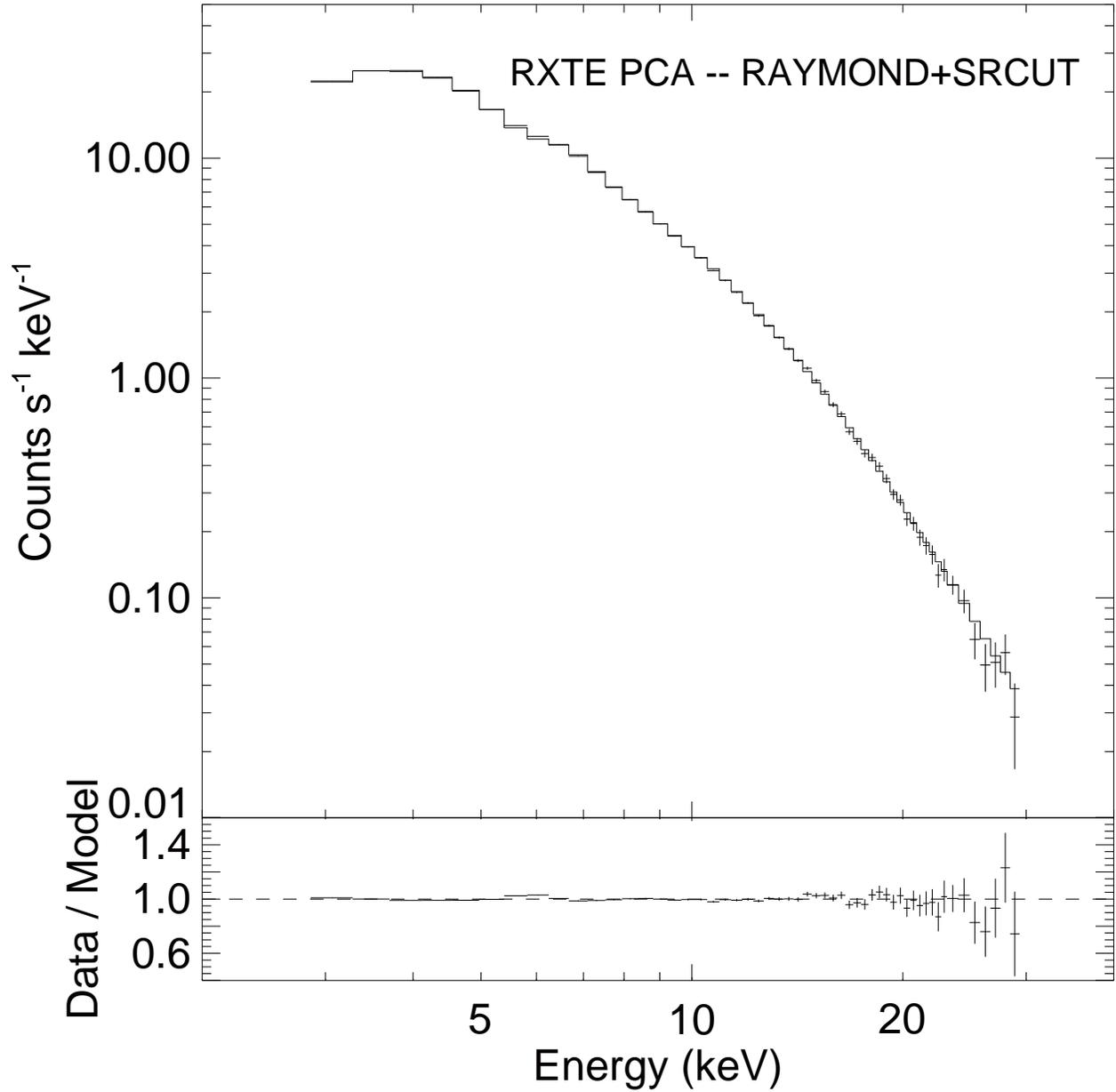}
\caption{Spectrum of G347.3$-$0.5 as observed by the {\it RXTE}~PCA
and fit using the {\it RAYMOND}+{\it SRCUT} model.\label{fig9}}
\end{figure}

\clearpage
\begin{figure}
\figurenum{10}
\epsscale{0.80}
\plotone{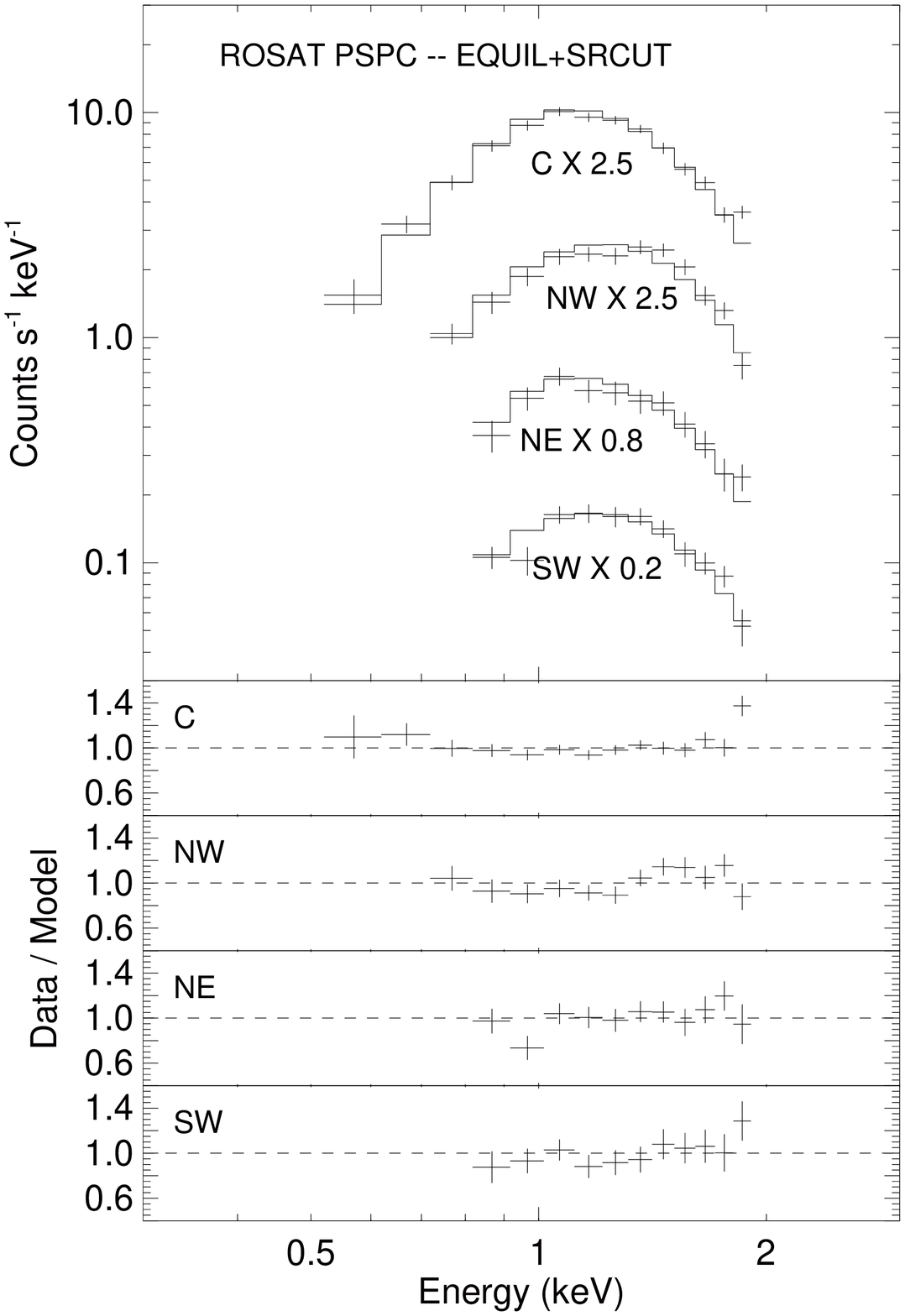}
\epsscale{1.00}
\caption{Spectra of the center (C), the northwestern rim (NW), the
northeastern rim (NE) and the southwestern rim (SW) of G347.3$-$0.5
as observed by the {\it ROSAT}~PSPC and fit using the
{\it EQUIL}+{\it SRCUT} model.\label{fig10}}
\end{figure}

\clearpage
\begin{figure}
\figurenum{11}
\plotone{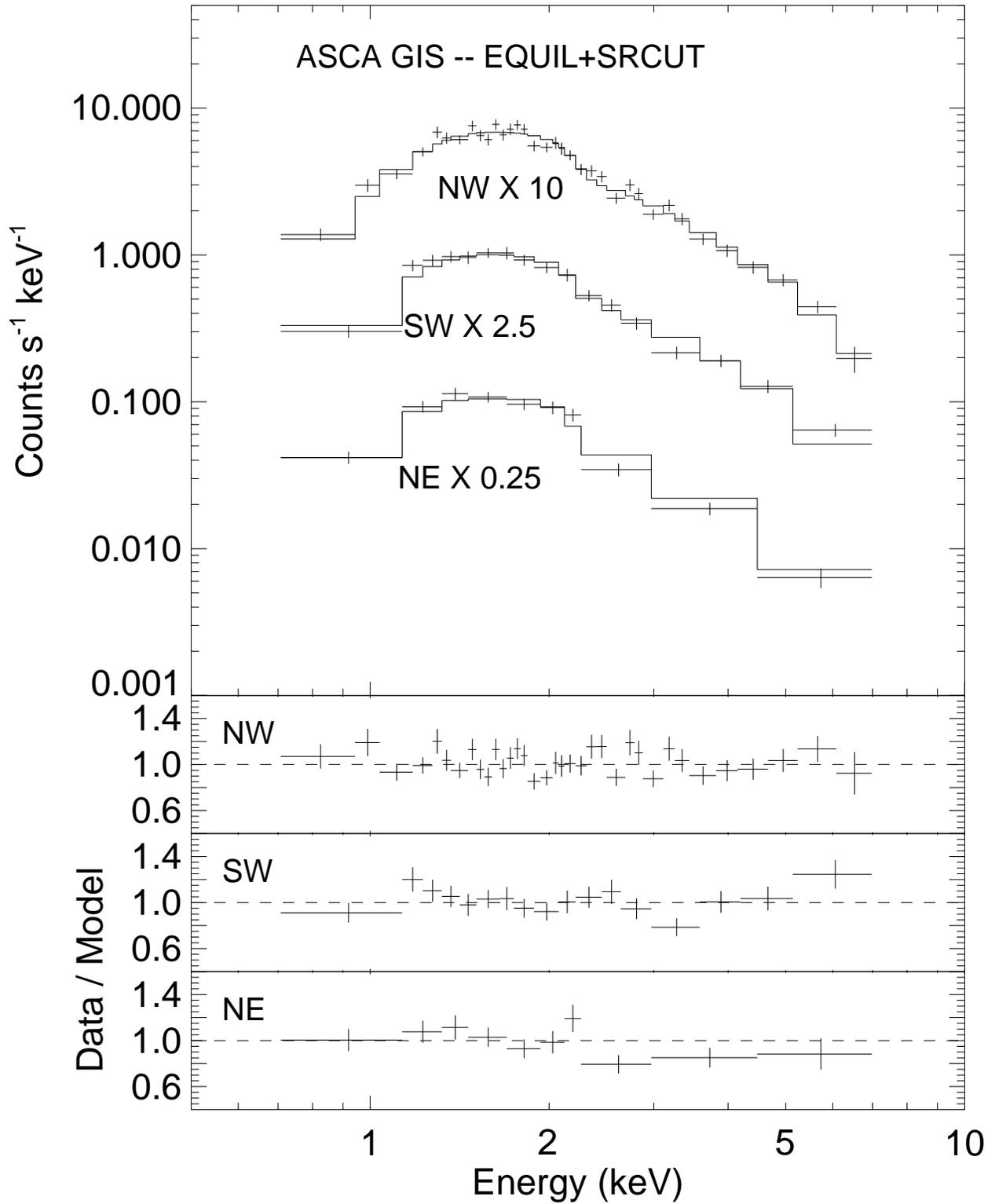}
\caption{Spectra of the northwestern rim (NW), the southwestern rim
(SW) and the northeastern (NE) rim of G347.3$-$0.5 as observed by the
{\it ASCA}~GIS and fit using the {\it EQUIL}+{\it SRCUT}
model.\label{fig11}}
\end{figure}

\clearpage
\begin{figure}
\figurenum{12}
\plotone{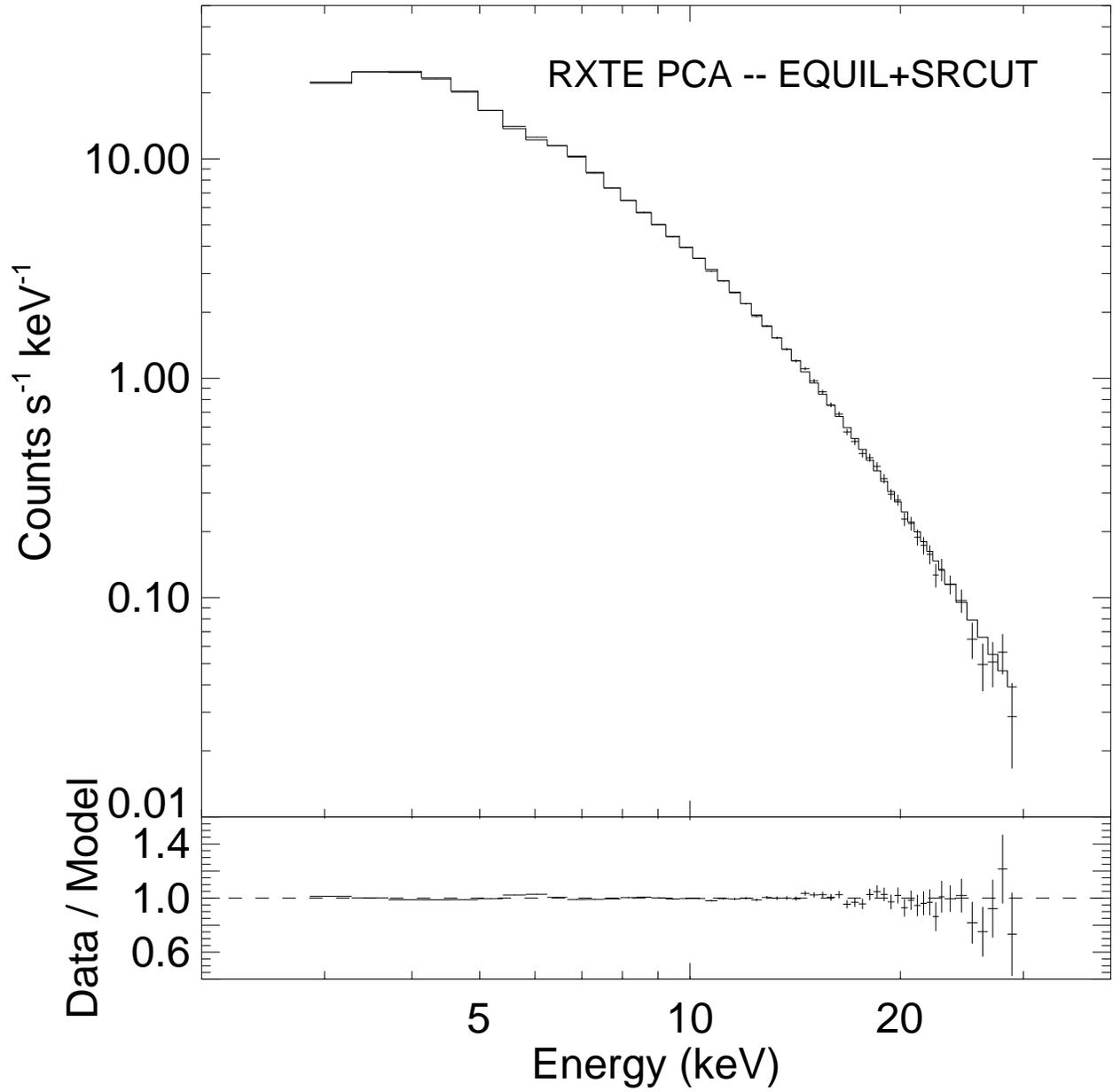}
\caption{Spectrum of G347.3$-$0.5 as observed by the {\it RXTE}~PCA
and fit using the {\it EQUIL}+{\it SRCUT} model.\label{fig12}}
\end{figure}

\clearpage
\begin{figure}
\figurenum{13}
\plotone{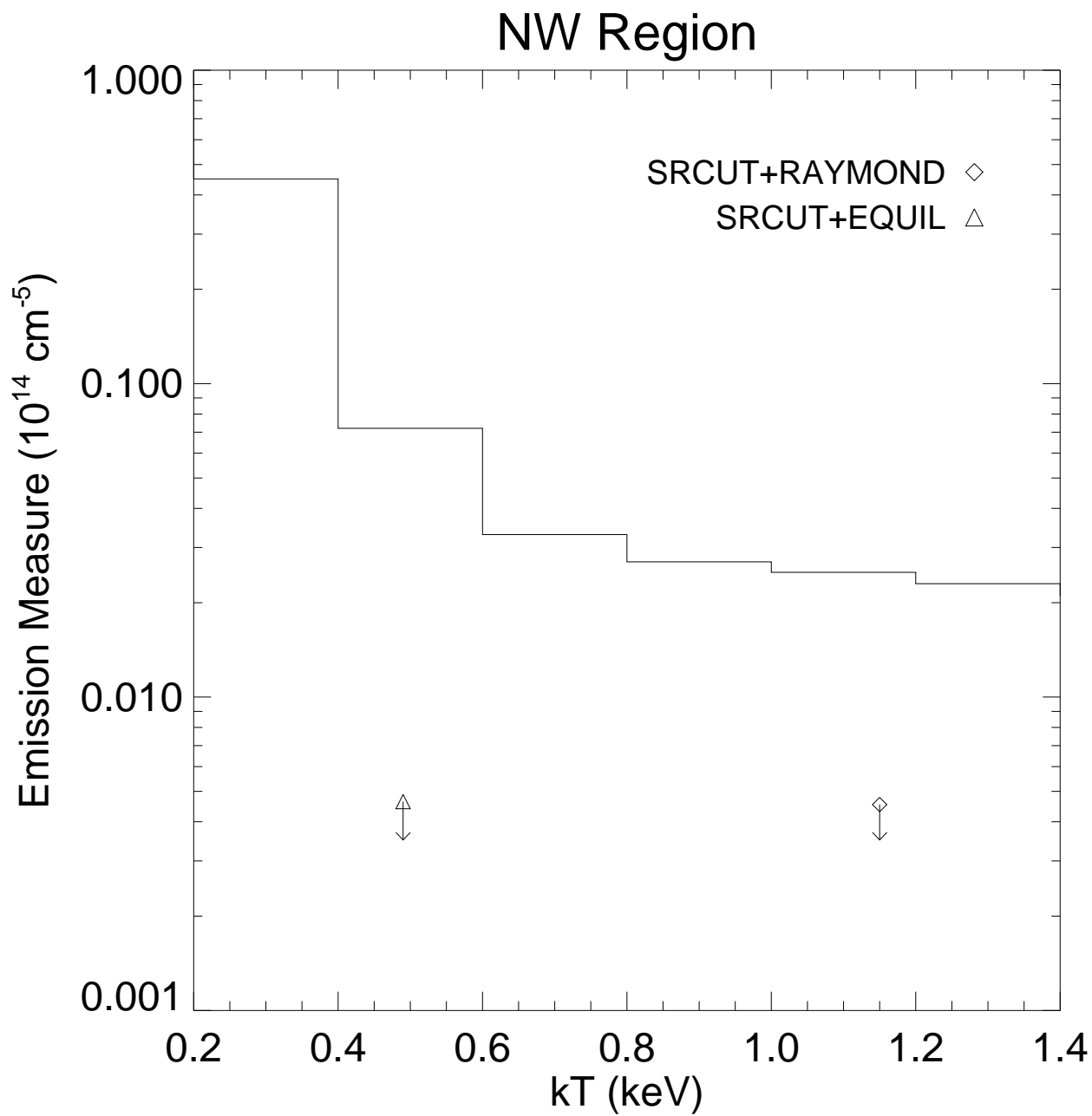}
\caption{Upper limits on Emission Measure for the northwestern
rim of G347.3$-$0.5, as calculated by \citet{S99}. The values 
(corresponding to 90\% confidence limits) for
the emission measure from our fits to the northwestern rim using
the $\it{RAYMOND}$+$\it{SRCUT}$ and $\it{EQUIL}$+$\it{SRCUT}$
are plotted using a diamond and a square, respectively.\label{fig13}}
\end{figure}

\clearpage 
\begin{figure}
\figurenum{14}
\plotone{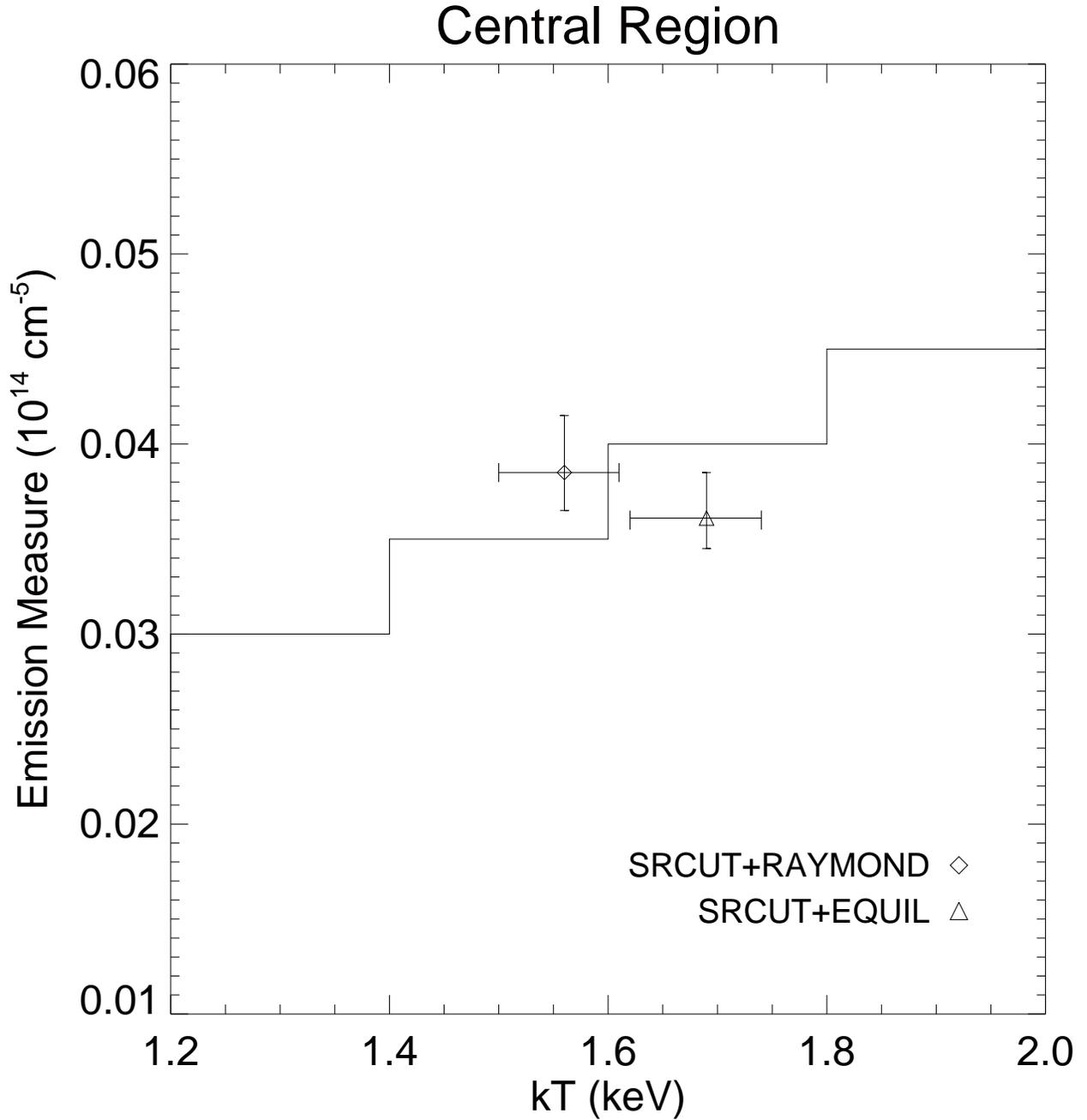}
\caption{Estimates of the Emission Measure for the central diffuse 
emission from G347.3$-$0.5, as calculated by \citet{S99}. The values 
(with 90\% confidence intervals) for
the emission measure from our fits to the northwestern rim using
the $\it{RAYMOND}$+$\it{SRCUT}$ and $\it{EQUIL}$+$\it{SRCUT}$
are plotted using a diamond and a square, respectively.\label{fig14}}
\end{figure}

\clearpage
\begin{figure}
\figurenum{15}
\plotone{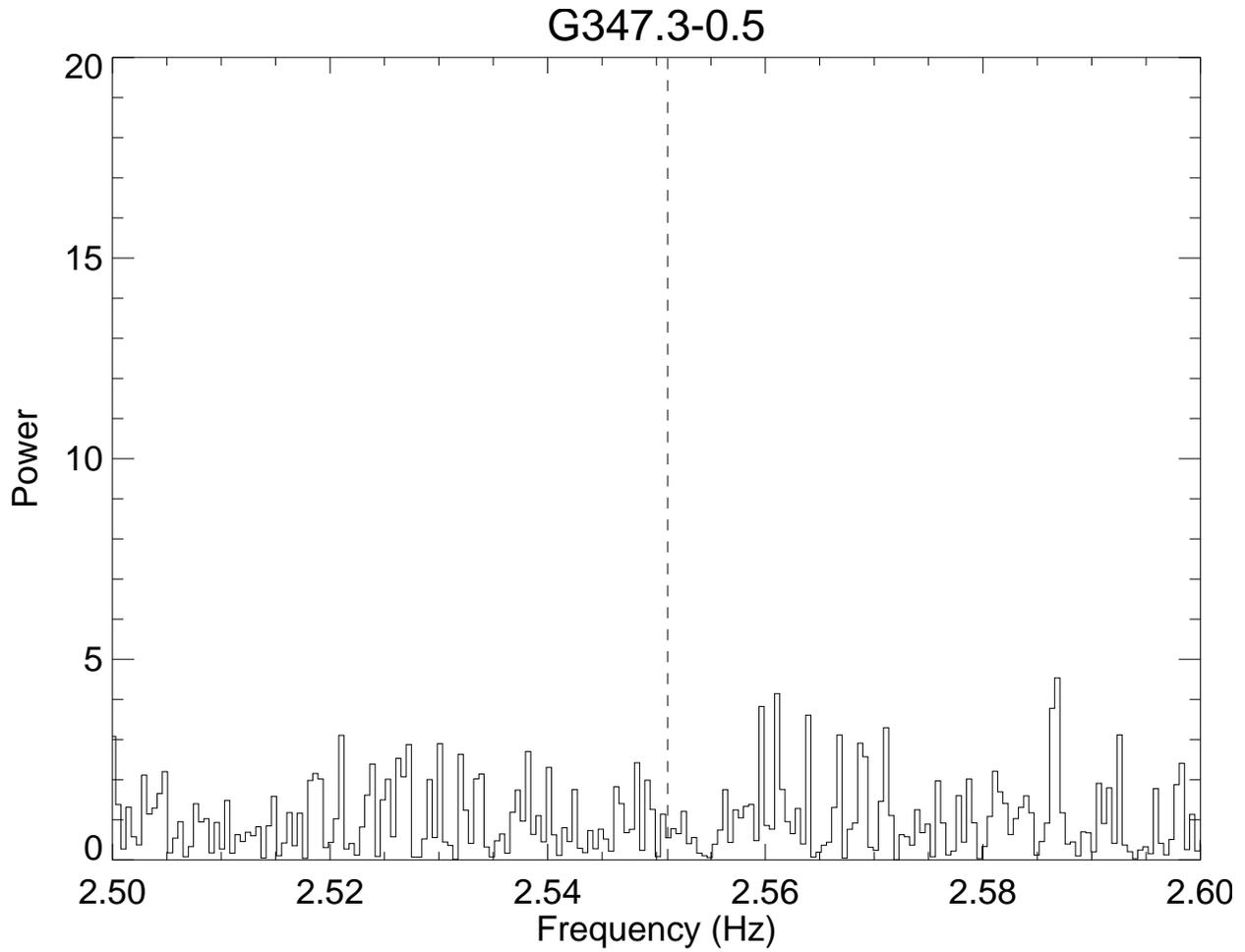}
\caption{Power spectrum of {\it RXTE}~PCA observation of G347.3$-$0.5.
The frequency corresponding to the putative pulsar (392 ms or 2.55 Hz)
is indicated by the vertical dashed line. No evidence for 
pulsations is detected at this frequency.\label{fig15}}
\end{figure}

\clearpage
\begin{figure}
\figurenum{16}
\plotone{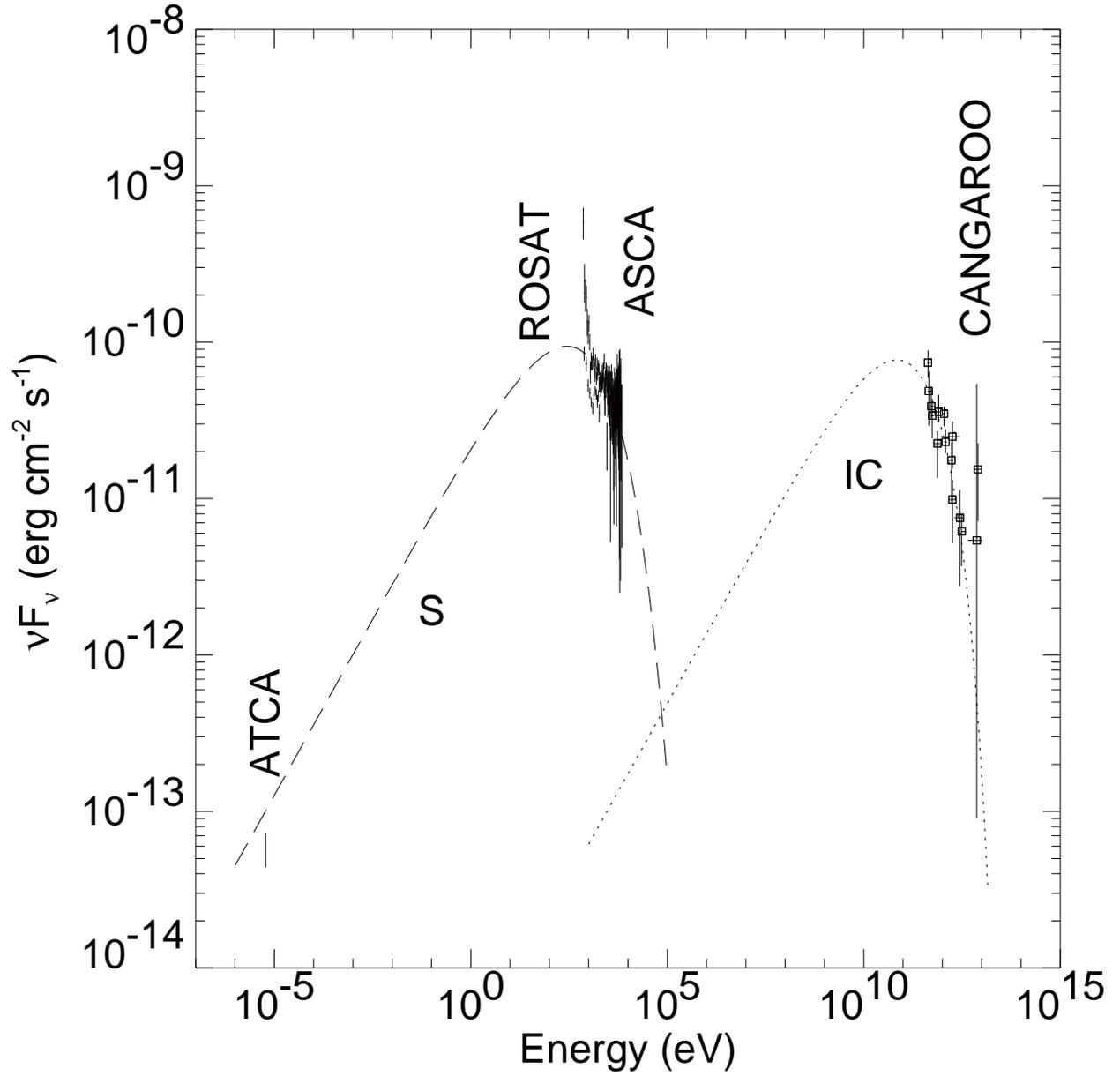}
\caption{Radio to gamma-ray photon energy-flux spectrum of
the northwestern rim of G347.3$-$0.5. See Section 
\ref{NWRimSection}.\label{fig16}}
\end{figure}

\clearpage

\begin{deluxetable}{lcccccc}
\tabletypesize{\scriptsize}
\tablecaption{{\it ROSAT}~PSPC, {\it ASCA}~GIS and {\it RXTE}~PCA
Observations of G347.3$-$0.5\label{ObsTable}}
\tablewidth{0pt}
\tablehead{ 
& & & \colhead{Observed} & & & \colhead{Exposure}\\
\colhead{Satellite and} & & \colhead{Observation} & \colhead{Portion of} 
& \colhead{R.A.} & \colhead{Decl.} & \colhead{Time}\\
\colhead{Instrument} & \colhead{Obs ID} & \colhead{Date} & 
\colhead{G347.3$-$0.5} & \colhead{(J2000.0)} & \colhead{(J2000.0)} 
& \colhead{(Seconds)} } 
\startdata
{\it ROSAT}~PSPC & RP500198N00 & 1992 September 23 & All 
& 17 13 33.6 & $-$39 48 36 & 2758 \\
{\it ASCA}~GIS & 55057000 & 1997 March 23 & NW Rim & 17 12 17.8 & $-$39 35 
31 & 5428 \\ 
{\it ASCA}~GIS & 55058000 & 1997 March 24 & SW Rim & 17 12 53.8 & $-$39 54 
23 & 5776 \\
{\it ASCA}~GIS & 55059000 & 1997 March 25 & NE Rim & 17 14 28.3 & $-$39 35 
26 & 3924 \\
{\it RXTE}~PCA & 40145-01 & 1999 June 15 & All & 17 14 13 & $-$39 50 30 & 
38371 \\
{\it RXTE}~PCA & 40145-02 & 1999 June 14 & Background \#1 & 17 22 26 & 
$-$37 18 10 & 8064 \\
{\it RXTE}~PCA & 40145-03 & 1999 June 14 & Background \#2 & 17 05 24 & 
$-$42 20 32 & 3229 \\
\enddata
\end{deluxetable}

\clearpage

%
%
\begin{deluxetable}{lcccc}
\tablecaption{Best-Fit Parameters -- $\it{SRCUT}$+$\it{RAYMOND}$
Model\tablenotemark{a}\label{SRCUTRAYFitTable}}
\tablewidth{0pt}
\tablehead{
& & & & \colhead{90\% Confidence}\\
\colhead{Region} & \colhead{Model} & \colhead{Parameter (Units)} &
\colhead{Value} & \colhead{Limits}
}
\startdata
NW & $\it{WABS}$ & $N$$_H$ (10$^{22}$ cm$^{-2}$) & 0.74 & 0.71, 0.77\\
& $\it{SRCUT}$ & $\alpha$ & 0.55 & 0.51, 0.59 \\
& $\it{SRCUT}$ & $\nu$$_{cutoff}$ (10$^{17}$ Hz) & 1.45 & 1.38, 1.59\\
& $\it{SRCUT}$ & Normalization $K$ (Jy) & 4.02 & 3.89, 4.09\\
& $\it{RAYMOND}$\tablenotemark{b} & $\it{kT}$ (keV) & 1.15 & \nodata\\
& $\it{RAYMOND}$ & Emission Measure $\it{EM}$\tablenotemark{c} &
$\leq$ 4.54 $\times$ 10$^{-3}$ & \nodata\\
& & (cm$^{-3}$)\\
NE & $\it{WABS}$ & $N$$_H$ (10$^{22}$ cm$^{-2}$) & 0.61 & 0.45, 0.76\\
& $\it{SRCUT}$ & $\alpha$ & 0.37 & 0.34, 0.52 \\
& $\it{SRCUT}$ & $\nu$$_{cutoff}$ (10$^{17}$ Hz) & 2.67 & 2.40, 2.96\\
& $\it{SRCUT}$ & Normalization (Jy) & 4.07 $\times$ 10$^{-2}$ & 
1.38 $\times$ 10$^{-2}$, \\
& & & & 0.11\\ 
& $\it{RAYMOND}$\tablenotemark{b} & $\it{kT}$ (keV) & 1.15 & 0.99,
1.71\\
& $\it{RAYMOND}$ & Emission Measure $\it{EM}$\tablenotemark{c} &
6.11 $\times$ 10$^{-3}$ & 3.05 $\times$ 10$^{-3}$, \\
& & (cm$^{-3}$) & & 9.69 $\times$ 10$^{-3}$\\
SW & $\it{WABS}$ & $N$$_H$ (10$^{22}$ cm$^{-2}$) & 0.63 & 0.60, 0.65\\
& $\it{SRCUT}$ & $\alpha$ & 0.50 & 0.48, 0.53 \\
& $\it{SRCUT}$ & $\nu$$_{cutoff}$ (10$^{17}$ Hz) & 2.43 & 2.27, 2.60\\
& $\it{SRCUT}$ & Normalization (Jy) & 0.72 & 0.43, 0.92 \\
& $\it{RAYMOND}$\tablenotemark{b} & $\it{kT}$ (keV) & 3.31 & 2.82, 4.00\\
& $\it{RAYMOND}$ & Emission Measure $\it{EM}$\tablenotemark{c} &
3.66 $\times$ 10$^{-3}$ & 2.63 $\times$ 10$^{-3}$, \\
& & (cm$^{-3}$) & & 4.78 $\times$ 10$^{-3}$\\
C & $\it{WABS}$ & $N$$_H$ (10$^{22}$ cm$^{-2}$) & 0.49 & 0.47, 0.51\\
& $\it{SRCUT}$ & $\alpha$ & 0.48 & 0.47, 0.51 \\
& $\it{SRCUT}$ & $\nu$$_{cutoff}$ (10$^{17}$ Hz) & 2.41 & 2.29, 2.52 \\
& $\it{SRCUT}$ & Normalization (Jy) & 1.28 & 0.98, 1.30 \\
& $\it{RAYMOND}$\tablenotemark{b} & $\it{kT}$ (keV) & 1.56 & 1.50, 1.61\\
& $\it{RAYMOND}$ & Emission Measure $\it{EM}$\tablenotemark{c} &
3.85 $\times$ 10$^{-2}$ & 3.65 $\times$ 10$^{-2}$, \\
& & (cm$^{-3}$) & & 4.15 $\times$ 10$^{-2}$\\
Background\tablenotemark{d} & $\it{WABS}$ & $N$$_H$ (10$^{22}$ cm$^{-2}$)
& 1.8\tablenotemark{d} & \nodata \\
& $\it{POWER~LAW}$ & Photon Index & 1.8\tablenotemark{d} & \nodata \\
& $\it{POWER~LAW}$ & Normalization & 2.72 $\times$ 10$^{-3}$
& 2.57 $\times$ 10$^{-3}$, \\
& & (Photons keV$^{-1}$ cm$^{-2}$ & & 2.84 $\times$ 10$^{-3}$\\
& & sec$^{-1}$ at 1 keV) \\
& $\it{RAYMOND}$\tablenotemark{b} & $\it{kT}$ (keV) &
2.9\tablenotemark{d} & \nodata \\
& $\it{RAYMOND}$ & Emission Measure $\it{EM}$\tablenotemark{c} &
3.46 $\times$ 10$^{-2}$ & 3.38 $\times$ 10$^{-2}$, \\
& & (cm$^{-3}$) & & 3.55 $\times$ 10$^{-2}$
\enddata
\tablenotetext{a}{$\chi$$^2$/Degrees of Freedom = 808.25/471 = 1.72
using 501 PHA bins.}
\tablenotetext{b}{Abundance frozen at 1.0 and redshift frozen at 0.0.}
\tablenotetext{c}{See Section 3.}
\tablenotetext{d}{Galactic diffuse background as observed by
the $\it{RXTE}$ PCA over the energy range of 2 through 30 keV. $N$$_H$,
photon index and $\it{kT}$ have all been frozen to the values measured
by \citet{VM98}.}  
\end{deluxetable}

\clearpage

\begin{deluxetable}{lcccc}
\tablecaption{Best-Fit Parameters -- $\it{SRCUT}$+$\it{EQUIL}$
Model\tablenotemark{a}\label{SRCUTEQUFitTable}}
\tablewidth{0pt} 
\tablehead{
& & & & \colhead{90\% Confidence}\\
\colhead{Region} & \colhead{Model} & \colhead{Parameter (Units)} &
\colhead{Value} & \colhead{Limits}
}
\startdata
NW & $\it{WABS}$ & $N$$_H$ (10$^{22}$ cm$^{-2}$) & 0.74 & 0.70, 0.79 \\
& $\it{SRCUT}$ & $\alpha$ & 0.55 & 0.49, 0.57 \\
& $\it{SRCUT}$ & $\nu$$_{cutoff}$ (10$^{17}$ Hz) & 1.47 & 1.31, 1.48 \\
& $\it{SRCUT}$ & Normalization (Jy) & 4.48 & 2.70, 4.55 \\
& $\it{EQUIL}$\tablenotemark{b} & $\it{kT}$ (keV) & 0.49 & \nodata \\
& $\it{EQUIL}$ & Emission Measure $\it{EM}$\tablenotemark{c} 
& $\leq$4.64 $\times$ 10$^{-3}$ & \nodata \\
& & (cm$^{-3}$) & & \\
NE & $\it{WABS}$ & $N$$_H$ (10$^{22}$ cm$^{-2}$) & 0.66 & 0.56, 0.75 \\
& $\it{SRCUT}$ & $\alpha$ & 0.44 & 0.37, 0.50 \\
& $\it{SRCUT}$ & $\nu$$_{cutoff}$ (10$^{17}$ Hz) & 2.93 & 2.59, 3.48 \\
& $\it{SRCUT}$ & Normalization (Jy) & 0.14 & 0.03, 0.51 \\
& $\it{EQUIL}$\tablenotemark{b} & $\it{kT}$ (keV) & 1.34 & 1.08, 1.71 \\
& $\it{EQUIL}$ & Emission Measure $\it{EM}$\tablenotemark{c}
& 9.31 $\times$ 10$^{-3}$ & 6.56 $\times$ 10$^{-3}$, \\
& & (cm$^{-3}$) & & 1.31 $\times$ 10$^{-2}$ \\
SW & $\it{WABS}$ & $N$$_H$ (10$^{22}$ cm$^{-2}$) & 0.61 & 0.58, 0.69 \\
& $\it{SRCUT}$ & $\alpha$ & 0.52 & 0.49, 0.55 \\
& $\it{SRCUT}$ & $\nu$$_{cutoff}$ (10$^{17}$ Hz) & 2.47 & 2.31, 2.59 \\
& $\it{SRCUT}$ & Normalization (Jy) & 1.10 & 0.58, 1.74 \\
& $\it{EQUIL}$\tablenotemark{b} & $\it{kT}$ (keV) & 1.31 & \nodata \\
& $\it{EQUIL}$ & Emission Measure $\it{EM}$\tablenotemark{c}
& $\leq$2.52 $\times$ 10$^{-3}$ & \nodata \\
& & (cm$^{-3}$) & & \\
C & $\it{WABS}$ & $N$$_H$ (10$^{22}$ cm$^{-2}$) & 0.49 & 0.48, 0.51 \\
& $\it{SRCUT}$ & $\alpha$ & 0.50 & 0.49, 0.52 \\
& $\it{SRCUT}$ & $\nu$$_{cutoff}$ (10$^{17}$ Hz) & 2.44 & 2.34, 2.53 \\
& $\it{SRCUT}$ & Normalization (Jy) & 1.94 & 1.42, 2.61 \\
& $\it{EQUIL}$\tablenotemark{b} & $\it{kT}$ (keV) & 1.69 & 1.62, 1.74 \\
& $\it{EQUIL}$ & Emission Measure $\it{EM}$\tablenotemark{c}
& 3.61 $\times$ 10$^{-2}$ & 3.45 $\times$ 10$^{-2}$, \\
& & (cm$^{-3}$) & & 3.82 $\times$ 10$^{-2}$ \\
Background\tablenotemark{d} & $\it{WABS}$ & $N$$_H$ (10$^{22}$ cm$^{-2}$)
& 1.8\tablenotemark{d} & \nodata \\
& $\it{POWER~LAW}$ & Photon Index & 1.8\tablenotemark{d} & \nodata \\
& $\it{POWER~LAW}$ & Normalization & 3.07 $\times$ 10$^{-3}$ &
2.91 $\times$ 10$^{-3}$,\\ 
& & (Photons keV$^{-1}$ cm$^{-2}$ & & 3.18 $\times$ 10$^{-3}$\\
& & sec$^{-1}$ at 1 keV) \\
& $\it{RAYMOND}$\tablenotemark{b} & $\it{kT}$ (keV) & 
2.9\tablenotemark{d} & \nodata \\
& $\it{RAYMOND}$ & Emission Measure $\it{EM}$\tablenotemark{c}
& 3.11 $\times$ 10$^{-2}$ & 3.04 $\times$ 10$^{-2}$, \\
& & (cm$^{-3}$) & & 3.21 $\times$ 10$^{-2}$
\enddata
\tablenotetext{a}{$\chi$$^2$/Degrees of Freedom = 843.76/472 = 1.79
using 501 PHA bins.}
\tablenotetext{b}{Abundance frozen at 1.0 and redshift frozen at 0.0.}
\tablenotetext{c}{See Section 3.}
\tablenotetext{d}{Galactic diffuse background as observed by
the $\it{RXTE}$ PCA over the energy range of 2 through 30 keV. $N$$_H$,
photon index and $\it{kT}$ have all been frozen to the values measured
by \citet{VM98}.}
\end{deluxetable}

\clearpage

\begin{table}
\begin{center}
\caption{Derived Values for $E$$_{cutoff}$, $\Gamma$, $n$$_H$ and 
$n$$_e$ Based on Model Fit Parameters\label{DerivedFitTable}}
\begin{tabular}{lcccc}
\tableline
\tableline
& \multicolumn{2}{c}{$\it{SRCUT+RAYMOND}$}  
& \multicolumn{2}{c}{$\it{SRCUT+EQUIL}$}\\
Parameter& NW & C & NW & C\\
\tableline
$E$$_{cutoff}$ (TeV)\tablenotemark{a} & 19.6 & 25.3 & 19.7 & 25.4 \\
$\Gamma$\tablenotemark{b} & 2.10 & 1.96 & 2.10 & 2.00 \\
$n$$_H$ (cm$^{-3}$) & $\leq$1.86 $\times$ 10$^{-2}$ & 0.054 &
$\leq$1.89 $\times$ 10$^{-2}$ & 0.052 \\
$n$$_e$ (cm$^{-3}$) & $\leq$2.23 $\times$ 10$^{-2}$ & 0.065 &
$\leq$2.26 $\times$ 10$^{-2}$ & 0.062 \\
\tableline
\end{tabular}
\tablenotetext{a}{Assuming a magnetic field of 10$\mu$G.}
\tablenotetext{b}{$\Gamma$ = 2$\alpha$ + 1.}
\end{center}
\end{table}

\clearpage 

\begin{deluxetable}{lcccc}
\tablecaption{Line Energies and Normalizations for Emission Feature
Detected Near 6.4 keV in $\it{RXTE}$~PCA Spectra\label{RXTEPCATable}}
\tablewidth{0pt}
\tablehead{
& \colhead{Line} & \colhead{90\% Confidence} & 
& \colhead{90\% Confidence}\\
\colhead{Pointing} & \colhead{Energy (keV)} & \colhead{Limits}
& \colhead{Normalization\tablenotemark{a}} & 
\colhead{Limits\tablenotemark{a}}
}
\startdata
G347.3$-$0.5 & 6.41 & 6.38, 6.42 & 5.05 & 4.70, 5.40 \\
Background \#1 & 6.51 & 6.47, 6.56 & 4.80 & 4.11, 5.59 \\
Background \#2 & 6.26 & 6.13, 6.37 & 7.24 & 5.48, 9.91 \\
Joint Fit\tablenotemark{b} & 6.43 & 6.40, 6.45 & 5.06 & 4.76, 5.38 \\
Cas A\tablenotemark{c} & 6.66 & 6.64, 6.67 & 58.4 & 57.8, 59.2 
\enddata
\tablenotetext{a}{In units of 10$^{-4}$ photons cm$^{-2}$ sec$^{-1}$.}
\tablenotetext{b}{Simultaneous fitting to the $\it{RXTE}$~PCA spectra
of G347.3$-$0.5 and both of the background pointings.}
\tablenotetext{c}{See Section \ref{IronFeatureSubSection}.}
\end{deluxetable}

\clearpage

\begin{deluxetable}{lcccccc}
\tabletypesize{\scriptsize}
\tablecaption{Properties of Shell-Type SNRs with Non-Thermal X-ray
Emission\label{ShellSNRsTable}}
\tablewidth{0pt}
\tablehead{
\colhead{Property} & \colhead{SN 1006} & \colhead{Reference} &
\colhead{G266.2$-$1.2} & \colhead{Reference} & \colhead{G347.3$-$0.5}
& \colhead{Reference}
}
\startdata
Distance (kpc) & 2.2 & (1) & 1-2 & (2) & 6 & (3), (4) \\
Ambient Density $N$ (cm$^{-3}$) & $\approx$ 0.1 & (5) & 
0.05\tablenotemark{a} & (2) & $\approx$0.05-0.07 & (6) \\
Angular Size (arcmin) & 30 & (7) & 120 & (7) & 65 $\times$ 55 & (7)\\
Dynamical Age (yr) & 996 & (1) & $\approx$14000\tablenotemark{a} & (2) & 
$\approx$8000 & (4) \\
Flux Density $S_{20}$ (Jy) & 19 & (7) & 50(?) & (7) & 4 & (4) \\
$\alpha$ & 0.6 & (7) & 0.3(?) & (7) & $\approx$0.55 & (6) \\
$B$ ($\mu$G) & 40 & (5) & \nodata & -- & 150$_{-80}^{+250}$ & (6)\\
$B$-field Direction & Radial & (8) & Tangential & (9) & \nodata & -- \\
Progenitor Type & Ia & (1) & II & (2) & II & (3), (4) 
\enddata
\tablecomments{References: (1) \citet{Winkler02}, (2) \citet{Slane01}, 
(3) \citet{S99}, (4) \citet{ESG01}, (5) \citet{A01}, (6) This paper, 
(7) \citet{Green01}, (8) \citet{Reynolds93} and (9) \citet{Combi99}.} 
\tablenotetext{a}{Calculated from expressions provided by \citet{Slane01}
and assuming a mean distance of 1.5 kpc, a volume filling factor for a
sphere of one-quarter and an explosion kinetic energy of 10$^{51}$ ergs.}
\end{deluxetable}

\end{document}